\documentclass[prd,aps,showpacs,nofootinbib,nobibnotes]{revtex4}
\usepackage{graphicx}
\usepackage{epstopdf}
\usepackage{epsfig,psfrag}
\usepackage{amsmath}

\def\gsim{\;\rlap{\lower 2.5pt
 \hbox{$\sim$}}\raise 1.5pt\hbox{$>$}\;}
\def\lsim{\;\rlap{\lower 2.5pt
   \hbox{$\sim$}}\raise 1.5pt\hbox{$<$}\;}

\begin{document}

\title{Galaxy Peculiar Velocities From Large-Scale Supernova Surveys as a Dark Energy Probe}

\author{Suman Bhattacharya}
\affiliation{ T-2, Theoretical Division, Los Alamos National Laboratory, Los Alamos, NM 87545}
\email{sumanb@lanl.gov}

\author{Arthur Kosowsky, Jeffrey A. Newman, and Andrew R. Zentner}
\affiliation{Department of Physics and Astronomy, University of Pittsburgh, Pittsburgh, PA 15260} 

\preprint{LA-UR-09.4568}

\begin{abstract}
 
Upcoming imaging surveys such as the Large Synoptic Survey Telescope 
will repeatedly scan large areas of sky and have the potential to yield million-supernova catalogs. 
Type Ia supernovae are excellent standard candles and will provide distance measures 
that suffice to detect mean pairwise velocities of their host galaxies.  
We show that when combining these distance measures with
photometric redshifts for either the supernovae or their host galaxies, the mean 
pairwise velocities of the host galaxies will provide a dark energy probe which
is competitive with other widely discussed methods.  
Adding information from this test to type Ia supernova photometric luminosity
distances from the same experiment, plus the cosmic microwave background
power spectrum from the Planck satellite, improves the Dark Energy Task Force Figure of Merit
by a factor of 1.8. Pairwise velocity measurements require no 
additional observational effort beyond that required to 
perform the traditional supernova luminosity distance test, but may provide complementary 
constraints on dark energy parameters and the nature of gravity. Incorporating additional
spectroscopic redshift follow-up observations could provide important dark energy constraints
from pairwise velocities alone. Mean pairwise velocities are much less sensitive
to systematic redshift errors than the luminosity distance test or weak lensing techniques,
and also are only mildly affected by systematic evolution of supernova luminosity. 

\end{abstract}

%\keywords{Cosmology: theory, Cosmology: peculiar velocities, galaxies, standard candles: supernova, type Ia, statistics}
\maketitle

\section{Introduction}
\label{sec: intro}

Measurements of the distance--redshift relation of type Ia supernovae (SNeIa) 
firmly established contemporary accelerated cosmological expansion \citep{riess98, perlmutter98} 
and SNIa distances remain one of the most promising probes of dark energy.  
To determine whether the accelerated cosmological 
expansion is caused by an ubiquitous dark energy or 
large-scale deviations from general relativity, it is 
necessary to measure both the expansion of the universe 
and the dynamics of the structure formation 
\citep{zhang_etal05,linder05,zhan_knox06,wang_etal07,huterer_linder07,
linder_cahn08,zhan_etal08,mortonson_etal08,zhang08,hearin_zentner09,zhao_etal10, kb09}.  
The SNIa luminosity distance test provides information about 
the expansion rate of the universe, but does not provide information on structure 
formation (though SNIa magnifications may achieve this in the future, 
see Refs.~\citep{metcalf99,dodelson_vallinotto06,zentner_bhattacharya09}).  
Peculiar velocities are related to densities through a continuity equation, 
so peculiar velocity statistics provide one avenue to study the 
growth of cosmic structure (e.g., \citep{fabian07}).  
The most well-explored option for probing the peculiar velocity field 
is via redshift-space distortions imprinted on the galaxy power spectrum 
(e.g., \citep{linder08,white08,percival08}).  Peculiar velocities may be detectable 
with future microwave experiments via the kinetic Sunyaev-Zeldovich effect 
\citep{bk06,bk07,bk08} and from large samples of SNeIa with spectroscopic redshifts 
\citep{SNvel}. In this paper, we examine the possibility of utilizing the 
mean pairwise velocity statistic, measured from SNeIa in a large 
photometric survey, to constrain dark energy.

Two well-studied statistics derivable from a sample of line-of-sight peculiar 
velocities are the {\em velocity correlation function} and 
the {\em mean pairwise velocity} \citep{ferreira99, sheth01}.  
The former is a two-point statistic expressing correlations in 
the peculiar velocities of objects are as a function of their separation.  
The mean pairwise velocity is a measure of 
the typical relative velocity of objects at a given separation.  
Peculiar velocities are sensitive to both the rate of structure growth in 
the universe and the rate of expansion of the universe.  Therefore, 
peculiar velocity measurements on cosmological scales may constrain the 
dark energy that drives cosmological acceleration and quenches late-time 
structure growth.

Traditionally, the bulk flow velocity has been measured by coupling measured galaxy 
redshifts with local distance indicators 
such as the fundamental plane of early-type galaxies \citep{feldman03,sarkar07}, 
the Tully-Fisher relation \citep{courteau00,borgani00,borgani00b}, 
or surface brightness fluctuations \citep{blakeslee00}.  
More recent studies \citep{watkins07, feldman08, feldman09} have measured 
significant bulk flows on scales of $100$~Mpc.
Radial velocity measurements have been used 
to reconstruct the velocity and density fields \citep{dekel99}.  
Reconstruction methods provide a way to test the gravitational 
instability theory and to measure the bias between the galaxy and mass density fields.  
Such studies are limited to the relatively local Hubble flow ($z \lesssim 0.1$), 
primarily because a constant fractional error in distance corresponds to a larger 
velocity interval at higher redshifts, an error that eventually overcomes the signal.

Type Ia supernovae, in contrast, are well-calibrated standard candles, and 
at cosmological distances SNeIa are more 
reliable distance indicators than those 
previously used for measuring peculiar velocities.  
Indeed, the dipole and quadrupole moments of the local bulk flow 
velocity have been measured to higher precision with the current 
data set of a few hundred SNeIa than with reconstructions based 
upon catalogs of many thousands of galaxies \citep{SNvel}.

SNeIa that are physically near each other exhibit coherent 
motion as they are influenced by correlated 
density structures.  
Therefore, the errors in the luminosity distance measurements 
of pairs of SNeIa should be correlated at low redshift ($z \lesssim 0.1$).  
Ignoring this correlation can lead to systematic 
biases in the determination of dark energy parameters \citep{SNLSvel,hui_greene06,cooray_sn06}.  
Alternatively, one can treat these correlated shifts in luminosity distance as ``signal,'' 
because peculiar velocities depend upon cosmological parameters.  
This signal has led to useful, independent constraints on the low-redshift normalization of the 
matter power spectrum, $\sigma_8$, and the total matter density, $\Omega_m$ \citep{SNvel107,SNvel07}.  
Unfortunately, direct measurements of 
the velocity correlation remain limited to relatively low redshifts.  
Even in an optimistic scenario of measurements of one million SNeIa, 
all with full spectroscopic follow-up, 
the velocity correlation can only 
be measured to a redshift of $z \simeq 0.5$ \citep{zhang08}.

In contrast to peculiar velocity correlation measurements, mean pairwise velocity is 
a linear statistic so its errors vary more mildly with redshift.  
In this study, we show that it will be possible to obtain interesting cosmological
information from mean pairwise velocities
to a redshift of $z=0.9$ in a large photometric survey of SNeIa, such as that 
planned for the Large Synoptic Survey Telescope (LSST) which is anticipated
to increase dramatically our current catalog of SNe1a, by a factor of nearly 1000. 
We demonstrate that 
such a measurement can provide dark energy constraints
that complement luminosity distance measurements 
under optimistic, but reasonable, assumptions.  
The constraints from mean pairwise velocities are 
also useful because they may be estimated with relatively little additional 
observational effort beyond that already required to use SNeIa 
to map luminosity distance or to detect cosmic lensing magnification \citep{zentner_bhattacharya09}. 
We find that combining the mean pairwise velocity measurements with 
distance measurements of SNeIa will sharpen constraints on the dark energy 
parameters compared to those inferred from luminosity distances alone.  
In particular, mean pairwise velocity constraints can improve 
the dark energy Figure of Merit from SNeIa as defined by the Dark Energy Task Force \citep{detf} 
(DETF) by a factor of 1.8. We additionally demonstrate that mean pairwise velocities, being
a differential statistic,
are potentially much less sensitive to systematic errors than other commonly considered
observational techniques. 
Ultimately, this property may make mean pairwise velocities one of the
most practically useful probes of dark energy.

Following the DETF, we describe the dark energy 
in terms of three phenomenological parameters: its current energy 
density $\Omega_\Lambda$ and two parameters describing the 
redshift evolution of its equation of state, $w_0$ and $w_a$, such that 
$w(a) = w_0 + (1-a)w_a$.  The additional cosmological parameters 
upon which the velocity field depends are the 
large-scale normalization of the matter power spectrum $\Delta_{\zeta}$, 
the power-law index of the primordial power spectrum $n_S$, 
the Hubble parameter $h$, the curvature of the universe $\Omega_k$, 
and the present-day matter density $\Omega_m$.  
In addition, we treat the photometric redshift (photo-z) dispersion, $\sigma_z$, 
as a free parameter with priors.  We label our set of parameters ${\bf p}$.  
We consider a fiducial cosmological model similar to the WMAP 5-year results \citep{WMAP5}:  
$\Delta_{\zeta}=2.0 \times 10^{-9}$, $n_S=0.95$, $h=0.71$, $\Omega_k=0$, $\Omega_m= 0.25$, 
$\Omega_\Lambda= 0.75$, $w_0= -1$, and $w_a= 0$.

The paper is organized as follows. In Section II, we describe our assumed 
SNIa survey specifications and review the estimation of 
supernova line-of-sight peculiar velocities from observed supernova brightnesses and redshifts.  
Section III describes a halo model calculation of the mean pairwise velocity as
a function of cosmological parameters.  
Sections IV and V quantify various sources of systematic and statistical errors 
that impact SNIa pairwise velocity measurements respectively.  
We present our results for dark energy parameter constraints in Section VI, using two
different sets of prior constraints. We also derive limits on systematic effects that must be
obtained to have the resulting parameter bias be smaller than the calculated statistical errors.  
In Section VII, we summarize the kinds of observational efforts required to meet the prospects
outlined in this paper, along with a brief discussion of the systematic error properties of
mean pairwise velocities compared to other dark energy probes. 

%%%%%%%%%%%%%%%%%%%%%%%%%%%%%%%%%%%%%%%%%%%%%%%%%%%%%%%%%%%%%%%%%%%%%%%%%%%%%%%%%%%%%%%%%%%
%%%%%%%%%%%%%%%%%%%%%%%%%%%%%%%%%%%%%%%%%%%%%%%%%%%%%%%%%%%%%%%%%%%%%%%%%%%%%%%%%%%%%%%%%%% 
\section{Large-Area Photometric Supernova Surveys}
\label{sec: survey_specs}

Forthcoming large-scale imaging surveys such as  
LSST or the Panoramic Survey Telescope and Rapid 
Response System (PanSTARRS) \citep{panstarrs,LSST,LSSTBook} will discover 
$10^4$ to $10^6$ SNeIa.  These SNeIa may be 
observed with broadband photometry with exposures 
spaced several days apart.  
To infer cosmological parameters from peculiar velocities, 
reliable distance measurements are needed.  These will likely be obtained from a 
well-characterized subset of the supernovae discovered by any survey.  The 
particular characteristics of this subset depend upon survey strategy and 
are difficult to anticipate.   

For ease of comparison with published studies, 
we adopt survey specifications similar to what may be 
achieved with a survey similar to LSST. 
We assume a total of $3\times 10^5$ SNeIa out to $z=1.2$, collected
over a dedicated supernova survey region of 300 square degrees; 
this corresponds to a SNIa surface density of 1000 deg$^{-2}$.  
This number density corresponds to 
the ``d2k'' survey described in \citep{zhan08} and 
such a dedicated survey may be undertaken as part of 
the science goals of the LSST \citep{LSSTBook}.  
We assume redshifts estimated using broadband photometry with 
a redshift-dependent, normally-distributed error 
of $\sigma_z=\sigma_{z0}(1+z)$.  DETF specifies an error range 
of $\sigma_{z0}=0.01$ for an optimistic scenario
to $\sigma_{z0}=0.05$ for a pessimistic scenario; 
in our parameter forecasts, we allow $\sigma_{z0}$ to vary along 
with the cosmological parameters.
Following \citet{zhan08}, we model the SNeIa redshift distribution as 
\begin{equation}
\frac{d^3n}{d\Omega\, dz\, dt} \propto  
\begin{cases}
\exp(3.12z^{2.1})-1, & \text{$z\le 0.5$,} \\ 
\left(\exp(3.12z^{2.1})-1\right)\exp(-12.2(z-0.5)^2), & \text{$z>0.5$.}
\end{cases}
\label{snrate}
\end{equation}

To the extent that SNeIa are standardizable candles, photometric observations
will yield a distance modulus $\mu$ and a luminosity distance $d_L$ via the usual relation
\begin{equation}
\mu = 2.17 \ln\left(\frac{d_L}{\rm Mpc}\right) + 25.
\label{mudef}
\end{equation}
The luminosity distance is obtained from the cosmological redshift $z$ via the definition
\begin{equation}
d_L(z) = (1+z) d_C(z) = (1+z)c\int_0^z \frac{dz'}{H(z')}  ,
\label{dLz}
\end{equation}
where $d_C(z)$ is the comoving line-of-sight distance to a galaxy at redshift $z$,
$H(z)$ is the Hubble parameter as a function of redshift, and a geometrically
flat universe has been assumed in the second equality. The evolution of the Hubble parameter, 
and thus the luminosity distance, depends on the assumed cosmological model. For a given
supernova, its measured redshift is the difference between its cosmological redshift
and the additional Doppler shift due to its line-of-sight velocity,
\begin{equation}
z_{\rm meas} = z(\mu) - \frac{v_{\rm los}}{c}(1+z(\mu)), 
\label{zmeas}
\end{equation}
where its cosmological redshift $z(\mu)$ can be obtained from its observed luminosity by inverting
Eqs.~(\ref{mudef}) and (\ref{dLz}). The factor of $(1+z)$ in Eq.~(\ref{zmeas}) accounts
for the cosmological redshift between the rest frame and the observation frame. 
For a given supernova with observed redshift and luminosity, its line-of-sight velocity can be
obtained by rearranging Eq.~(\ref{zmeas}) into
\begin{equation}
v_{\rm los} = \frac{cz(\mu) - cz_{\rm meas}}{1+z_{\rm meas}} ,
\label{vlos}
\end{equation}
where we have replaced $z(\mu)$ by $z_{\rm meas}$ in the denominator, which will always
be a good approximation for objects at cosmological distances where the first term in
Eq.~(\ref{zmeas}) is large compared to the second term. 

Traditional peculiar velocity estimates using other standard candles at cosmological
distances have been hampered by errors in distance estimates, which propagate
into errors in $z(\mu)$. For a galaxy with cosmological redshift $z=0.03$, 
a $10\%$ error in distance corresponds to an error in inferred cosmological redshift
equivalent to a peculiar velocity of 1000 km/s, with
the size of the error increasing proportional to redshift for $z \leq 1$. Large-area
supernova surveys offer two main advantages. First, supernovae are bright
enough and good enough standard candles
to provide convenient distance estimators out to $z=1$ and beyond. Second,
the anticipated large number of supernovae hold the promise of
determining average distances far more precisely than individual
distances, allowing precise determination of average velocity statistics
from large catalogs of supernovae. Of course, realizing this promise
requires controlling systematic errors in both distance and redshift
observations to a high level, so that averages over large ensembles of 
SNeIa reflect the actual velocity statistic. 
Both systematic and statistical errors will be considered following the
next Section, which outlines the application of the mean pairwise velocity statistic 
to supernova surveys.

%%%%%%%%%%%%%%%%%%%%%%%%%%%%%%%%%%%%%%%%%%%%%%%%%%%%%%%%%%%%%%%%%%%%%%%%%%%%%%%%%%%%%%%%%%%%%%%
%%%%%%%%%%%%%%%%%%%%%%%%%%%%%%%%%%%%%%%%%%%%%%%%%%%%%%%%%%%%%%%%%%%%%%%%%%%%%%%%%%%%%%%%%%%%%%%
\section{Mean Pairwise Peculiar Velocity}
\label{sec: theory}

The mean pairwise velocity $v(r,a)$ at a comoving separation $r$ and scale factor $a=1/(1+z)$ 
is the average over all pairs at a fixed comoving  separation of the relative peculiar
velocity of the two galaxies projected along the line joining them.  That is, 
\begin{equation}
v(r,a)= \frac{1}{N(r)}\sum_{i \ne j} ({\bf v}_i- {\bf v}_j) \cdot {\bf \hat r},
\label{eq:vij}
\end{equation}
where ${\bf v}_i$ is the peculiar velocity of supernova $i$ and 
$\bf {\hat r}$ is the unit vector in the direction of the separation of 
the two objects. 
The sum is over $N(r)$ pairs at a given comoving separation $r$.
(Note that the quantity which we write throughout this paper as ``$v(r,a)$'' 
is commonly written in the literature as ``$v_{ij}(r,a)$'' or ``$v_{12}(r,a)$.''   
We use this notation to avoid potential confusion with subscript labels 
for individual galaxies that we use below.)

The mean pairwise velocity for dark matter particles may be derived using the 
pair conservation equation \cite{davis77}.  However for galaxies, 
the pair conservation equation needs to be modified to account for 
evolution \cite{sheth01}. The resulting mean pairwise velocity for 
SNIa host galaxies with a comoving separation $r$ at a mean scale factor $a$ (assuming that
the redshift difference between the two galaxies corresponds to a scale factor difference much
smaller than $a$) can be written 
\begin{equation}
v(r,a)=-\frac{2}{3}H(a)a\frac{d \ln D_a}{d \ln a}b_{\rm gal}(a)\frac{r\bar{\xi}^{\rm dm}(r,a)}{1+\xi^{\rm gal}(r,a)} ,
\label{v12}
\end{equation}
where 
\begin{equation}
\xi^{\rm dm}(r,a)= \frac{D_a^2}{2\pi^2r} \int_0^{\infty} dk\,k \sin(kr) P(k) 
\end{equation} 
is the dark matter two-point correlation function, 
$P(k)$ is the dark matter power spectrum at wavenumber $k$, 
$H(a)$ is the Hubble parameter at a given redshift, and $D_a$ is the linear growth factor as a 
function of time, normalized so that $D_{a}=1$ at $z=0$. 
We also define the dark matter correlation function averaged over 
separations less than $r$ to be
\begin{equation}
\bar{\xi}^{\rm dm}(r,a) = \frac{3}{r^3}\int_0^r dr'\,r'^2 \xi^{\rm dm}(r',a).
\end{equation}
We are interested in the large-scale limit, so we model the correlation function of supernova host galaxies
using a deterministic linear bias relative to the dark matter $b_{\rm gal}(z)$, defined by
\begin{equation}
\xi^{\rm gal}(r,a)=b_{\rm gal}^2(z) \xi^{\rm dm}(r,a).  
\end{equation} 
The bias $b_{\rm gal}(z)$ in general varies with the galaxy separation, a variety of 
galaxy properties, and redshift \citep{galbias}.  In the large-scale limit, scale-independent 
bias is a fairly good assumption.  Following Ref.~\citep{zhan08}, 
we model $b_{\rm gal}(z)$ as $b_{\rm gal}(z)=1.0+0.6z$ 
to obtain the fiducial value of the galaxy bias as a function of redshift. 

With future photometric surveys potentially detecting more than a billion galaxies, we can expect that 
the correlation function of samples of galaxies matching the SN hosts can 
be measured to percent-level accuracy or better.  
Thus the uncertainty in $b_{\rm gal}$ will primarily 
be due to uncertainty in the cosmological 
parameters affecting the dark matter correlation function. We express $b_{\rm gal}$ in terms 
of the galaxy and predicted dark matter correlation functions, and use 
this bias value in Eq.~(\ref{v12}).

Only the line-of-sight component of the velocity can be obtained from
observations, while the mean pairwise velocity involves all three directional components of 
the velocity. We use the estimator for the mean pairwise velocity given a data set 
of line-of-sight velocities developed in Ref.~\citep{ferreira99}.  Consider two galaxies $i$ and $j$ at 
comoving positions ${\bf r}_i$ and ${\bf r}_j$ moving with peculiar velocities ${\bf v}_i$ and ${\bf v}_j$.  
The radial component of velocities can be written as ${v}^{r}_i= {\bf \hat r}_i\cdot{\bf v}_i$ and  
${v}^{r}_j= {\bf\hat r}_j\cdot{\bf v}_j$. 
Then an estimate for the pairwise velocity of the two galaxies $v_{ij}^{\rm est}$ is defined by
$\langle {v}^{r}_i-{v}^{r}_j\rangle = 
v_{ij}^{\rm est}{\bf\hat r}\cdot({\bf\hat r}_i+ {\bf\hat r}_j)/2$, where ${\hat r}$ is the 
unit vector along the line joining the two galaxies.  If we now consider a catalog of
line-of-sight galaxy velocities, minimizing $\chi^2$ between the actual pairwise velocities and
the estimate of the pairwise velocity at a given separation $r$ gives an estimator
for the pairwise velocity Eq.~(\ref{v12}) based on the catalog,
\begin{equation}
v^{\rm est}(r,a)= \frac{\sum_{\rm pairs}(v^{r}_i- v^{r}_j)p_{ij}}{\sum_{\rm pairs} p_{ij}^2} ,
\label{v_est}
\end{equation}
where the sums are over all pairs $i\neq j$ of galaxies at comoving separation $r$ and 
$p_{ij}= {\bf\hat r} \cdot ({\bf\hat r}_i + {\bf\hat r}_j)/2$. Note that this form for the projection
tensor $p_{ij}$ is applicable in the flat-sky limit and breaks down for large angular
separations; in particular it is zero if the two galaxies are in opposite sky directions. In this
paper, we consider a model supernova survey of 300 square degrees in a compact
sky region, and the $p_{ij}$ expression given here is always valid. To extend the results
here to a full-sky survey, or to survey patches which are separated by large angles,
a more complicated projection tensor must be used. The derivation is not conceptually 
difficult, but this will be deferred to future work giving more detailed estimates
of signal-to-noise ratios for particular observing strategies. 

Equation (\ref{v_est}) is a function of the separation $r$ between the two galaxies. To measure
this distance, we must use the estimated locations of each galaxy; this is subject to errors
which will be quantified in the next Section. The separation that is measured directly 
is the angle between two galaxies on the sky.  This angle can be converted to 
the transverse component of the distance between the two galaxies using the angular 
diameter distances corresponding to their redshifts. 

The expression in Eq.~(\ref{v_est}) is a very simple estimator which weights all pairs of velocities
uniformly. A more careful analysis of real data would use, for example, a signal-to-noise
weighting in the sum. This is not a major correction to the analysis in this paper, as we
limit the sums in Eq.~(\ref{v_est}) to pairs  with separations smaller than 100 Mpc; at larger
separations the signal becomes small.  
In principle, a signal-to-noise weighting can squeeze more information out of the data, using
pairs with larger separations, but it does not qualitatively change our results. Our estimator is
accurate, as we have shown explicitly in Fig.~4 of Ref.~\cite{bk07}, but suboptimal; an
optimal estimator will somewhat improve the constraining capability of a 
velocity survey compared to the analysis here, so our estimator is conservative. 

As we discuss further in Section~\ref{sec:sys_err} and Section~\ref{sec:results}, we 
can mitigate the influence of systematic redshift errors by considering a related projected statistic,
where the mean pairwise velocity is taken as a function of the angular separation of the
two galaxies rather than as a function of their three-dimensional separation.  
This is given by
\begin{equation}
{\tilde v}(\theta,a) = 
\int_0^{\pi_{\rm max}}  d\pi_t P(\pi_t | \theta, a) v(r, a) ,
\label{v_projected}
\end{equation}
where the line-of-sight comoving separation $\pi_t = d_C(a_2) - d_C(a_1)$ and $P(\pi_t | \theta, a)$ is the probability that a pair has line-of-sight separation $\pi_t$ given that it has
an angular separation on the sky $\theta$. We can write the three-dimensional separation $r$ in terms of the angular separation $\theta$ and $\pi_t$ as
\begin{equation}
r=\sqrt{\theta^2d_M(a)^2 + \pi_t^2},
\label{rdef}
\end{equation}
where 
\begin{equation}
d_M(a) = 
\begin{cases}
cH_0^{-1}\Omega_k^{-1/2} \sinh\left[ \Omega_k^{1/2}d_C(a) / (cH_0^{-1})\right], & \Omega_k > 0\\
d_C(a), & \Omega_k = 0\\
cH_0^{-1}|\Omega_k|^{-1/2} \sin\left[ |\Omega_k|^{1/2}d_C(a) / (cH_0^{-1})\right], & \Omega_k < 0
\end{cases}
\label{dMdef}
\end{equation}
is the transverse comoving distance to scale factor $a$; here $\Omega_k$ is the effective
curvature density, $\Omega_k = 1 - \Omega_m - \Omega_\Lambda$
(see Ref.~\cite{hog99} for a lucid discussion of various distance measures in cosmology).
If the redshift difference is small compared to unity,
$\pi_t H(z_1) \approx c(z_2 - z_1)$, though we compute the separation 
in full for all pairs. For a spatially flat universe, $d_M(a) = d_C(a)$. In our case,
we always consider separations with $r\ll cH_0^{-1}$ since the signal is only
significant on these scales. We therefore
always have $d_M(a) \approx d_C(a)$ to good accuracy, and for simplicity
we make this assumption throughout the rest of the paper and use comoving distances entirely.
We consider pairs of galaxies with line-of-sight 
comoving separations up to a maximum value $\pi_{\rm max}$ (in practice, we will measure 
redshift-space rather than comoving separations; we consider the impact of this in Section~\ref{subsec:photoz_effect}).  
The probability of a pair having line-of-sight separation $\pi_t$ given that it has
an angular separation on the sky $\theta$ is
\begin{equation}
P(\pi_t | \theta, a) = \frac{1 + \xi^{\rm gal}(r, a)}
{\int_0^{\pi_{\rm max}} d\pi_t \left[ 1 + \xi^{\rm gal}(r, a)\right]}
\label{Ppitheta}
\end{equation}
for $\pi_t < \pi_{\rm max}$ and $P(\pi_t |\theta, a) = 0$ for $\pi_t > \pi_{\rm max}$.  

An estimator for ${\tilde v}(\theta,a)$
from line-of-sight velocity data
is easily obtained by substituting $v^{\rm est}(r,a)$ for $v(r,a)$ in Eq.~(\ref{v_projected}). 
To compare with data, we bin this statistic in angular separation and redshift, putting each
pair in the redshift bin corresponding to the mean photometric redshift of the 
two galaxies in the pair. In this manner all pairs are included regardless of binning; we have verified 
that our results remain similar when modest changes are made to projection and binning schemes.  
Note a correction for scatter in measured redshifts must also be included, as discussed
below in the following Section.

Changing the maximum separation $\pi_{\rm max}$ considered in Eq.~(\ref{v_projected}) 
will modify the signal-to-noise ratio in measuring the projected pairwise velocity.  
A larger $\pi_{\rm max}$ increases the total
number of pairs considered, but the signal-to-noise ratio for each
pair decreases at larger $\pi_t$ (as measurement errors remain approximately unchanged 
but signal strength decreases), so their contribution is small. For the purposes of this paper, 
we adopt a cutoff of $\pi_{\rm max}= 100$ Mpc, which 
captures the great majority of the pairwise velocity signal. As a test of this effect, we find that including 
pairs out to separations two times larger only changes the signal-to-noise in measuring the projected 
pairwise velocity by around 10\%. Based on this, we conclude that including data from pairs with
separations larger than 100 Mpc should give only minimal improvements in parameter constraints 
compared to those presented here. We also impose a miminum separation of 20 Mpc
on the pairs we consider, to eliminate any systematic errors related to nonlinear effects.
The mean pairwise velocity is a declining function as
separation increases from 20 Mpc to 100 Mpc, as shown in Fig.~\ref{fig: photoz}; at smaller
scales, it turns over and decreases in linear theory.

%%%%%%%%%%%%%%%%%%%%%%%%%%%%%%%%%%%%%%%%%%%%%%%%%%%%%%%%%%%%%%%%%%%%%%%%%%%%%%%%%%%%%%%%%%%%%%
\section{Systematic Errors}
\label{sec:sys_err}

\subsection{Photometric Redshift Errors}
\label{subsec:photoz_effect}

Large imaging surveys will detect so many galaxies that it will not be feasible to obtain 
spectroscopic redshifts for the vast majority. We must settle for photometric redshift estimates
determined from the fluxes measured in the various observed bands.  These photometric redshifts will
be less accurate than spectroscopic redshifts, and may have complex 
error distributions. Here we consider a measured redshift distribution
described by a Gaussian of standard deviation $\sigma_z$ centered at the true redshift of each 
object. We neglect a possible
photometric redshift bias for two reasons:  First, 
in realistic surveys this bias can be 
calibrated by comparison with a manageable number of
spectroscopic SNIa observations \citep{detf,zhan08,zentner_bhattacharya09}.  
We emphasize that we utilize a normal distribution for definiteness, 
but a well-calibrated error distribution is what is necessary to proceed; errors need not be Gaussian in practice.  
Second, the expected level of photometric redshift bias is likely to be a small effect 
\citep{pinto05,frieman09} 
compared to the systematic errors in estimating distances that we consider below.  
As a result, we do not explicitly carry a bias through in the equations below, but we will 
present a test of the impact of a bias in photometric redshifts in Sec.~\ref{sec:results}.

In contrast, the photo-z dispersion, $\sigma_z$, essentially smooths 
the estimated velocity distribution of the observed sample 
and propagates scatter into galaxy pair separations.  
The latter effect can cause not only a scatter in inferred cosmological
parameter values, but also a systematic shift, which we calculate here.

The mean pairwise velocity $v(r,a)$, given in Eq.~(\ref{v12}), assumes that 
the three-dimensional separation $r$ between the SNeIa or their host galaxy 
pairs are known accurately; however, there will be non-negligible errors 
in observed redshifts.  Our simple normal-error model for the distribution of the
photometric redshift $z_p$, given a true redshift $z$, is
\begin{equation}
P(z_p|z,\sigma_z)= \frac{1}{\sqrt{2\pi\sigma_z^2}}\exp[-(z-z_p)^2/(2\sigma_z^2)].
\label{photoz}
\end{equation}
We take the photo-z dispersion to be $\sigma_z=\sigma_{z0}(1+z)$ 
with $\sigma_{z0}$ ranging from 0.01 to 0.05 \citep{pinto05,detf,wang_etal07,zhan08,frieman09}.
We explore the sensitivity of our results to prior knowledge of $\sigma_{z0}$ in 
\S~\ref{sec:results}.

Using Eq.~(\ref{photoz}) and the expression $H(z_p)\pi_{\rm t}=c(z_{p2}-z_{p1})$ for the 
local Hubble expansion about each SNIa, where $z_{p2}-z_{p1}$ is the photometric redshift difference
between a pair of supernovae, we write the probability of obtaining the 
observed line-of-sight separation $\pi_{\rm obs}$ for a given, true comoving 
line-of-sight separation $\pi_{\rm t}$ as 
\begin{equation}
P(\pi_{\rm obs}|\pi_{\rm t},\sigma_\pi)= \frac{1}{\sqrt{2\pi\sigma_\pi^2}}\exp\left[-(\pi_{\rm obs}-\pi_{\rm t})^2/(2\sigma_\pi^2)\right] ,
\label{separation_los}
\end{equation} 
where $\sigma_\pi= \sqrt{2}c\sigma_z/H(z)$ \citep{frieman09}.  
We assume that the photometric redshifts $z_p$, although they include the effects of 
peculiar motions, give a better measurement of the galaxy line-of-sight separation than
the cosmological redshifts $z(\mu)$, 
which must be determined via a distance measurement with uncertainties on the order of 10\%;
hence the line-of-sight positions of SNeIa are estimated using $z_p$.
The factor $\sqrt{2}$ in relating $\sigma_\pi$ to $\sigma_z$ accounts for 
uncertainties in the positions of the two galaxies in a pair, 
which are added in quadrature.  
Combining Eqs.~(\ref{v12}), (\ref{v_projected}), and (\ref{separation_los}), 
we get an expression for the projected, mean pairwise velocity accounting for 
a significant dispersion in photometric redshifts, 
\begin{equation}
{\tilde v}(\theta,a|\sigma_\pi(a))= \int_0^{\pi_{\rm max}} d\pi_t \int_0^\infty d\pi_{\rm obs} 
P(\pi_t|\theta,a)P(\pi_{\rm obs}|\pi_{\rm t},\sigma_\pi(a))\,
v((\theta^2 d_C(a)^2 + \pi_t^2)^{1/2},a).
\label{v12perp}
\end{equation}
We propose using this statistic as a cosmological probe.  We consider only positive values of $\pi_t$, 
so we count each pair only once.  This remains true if in some cases (due to errors) $\pi_t$ scatters below 
zero (in which case the separation is positive when the two members of the pair are exchanged).  

Both Eq.~(\ref{separation_los}) and the expression for $\sigma_\pi$ are valid only when 
$|z_{p2}-z_{p1}|\ll 1$; however, we should always be in this limit.  The maximum true separation 
we consider, $\pi_{\rm max}=100$ Mpc, corresponds to a redshift difference ranging
from 0.024 to 0.042 as $z$ ranges from 0 to 1; photo-z errors 
will broaden the distribution of separations via a Gaussian kernel with 
dispersion $\sigma=\sqrt{2} \sigma_z=\sqrt{2}\sigma_{z0}(1+z)$, which gives 
$\sigma_z = 0.056$ at $z=1$ for $\sigma_{z0}=0.02$.  To the degree that the assumption of small 
$|z_{p2}-z_{p1}|$ is violated, the small distance error 
induced by this approximation remains negligible, as the pairwise velocity does 
not vary rapidly on any scales of interest.  
One additional caveat is that these relations hold only for sufficiently large angular separations, 
corresponding to comoving separations greater than approximately 5 Mpc, so that nonlinear effects due to velocities within 
gravitationally bound objects (``fingers of god'') are insignificant.  
%The calculations presented in this paper barely dip below this limit at the smallest 
%angles considered in our lowest-redshift bin (0.5 degrees corresponds to a 4 Mpc separation 
%at $z=0.1$), but otherwise are well outside of this regime.  

%%%%%%%%%%%%%%%%%%%%%%%%%%%%%%%
\begin{figure*}
  \begin{center}
    \begin{tabular}{cc}
      \resizebox{85mm}{!}{\includegraphics{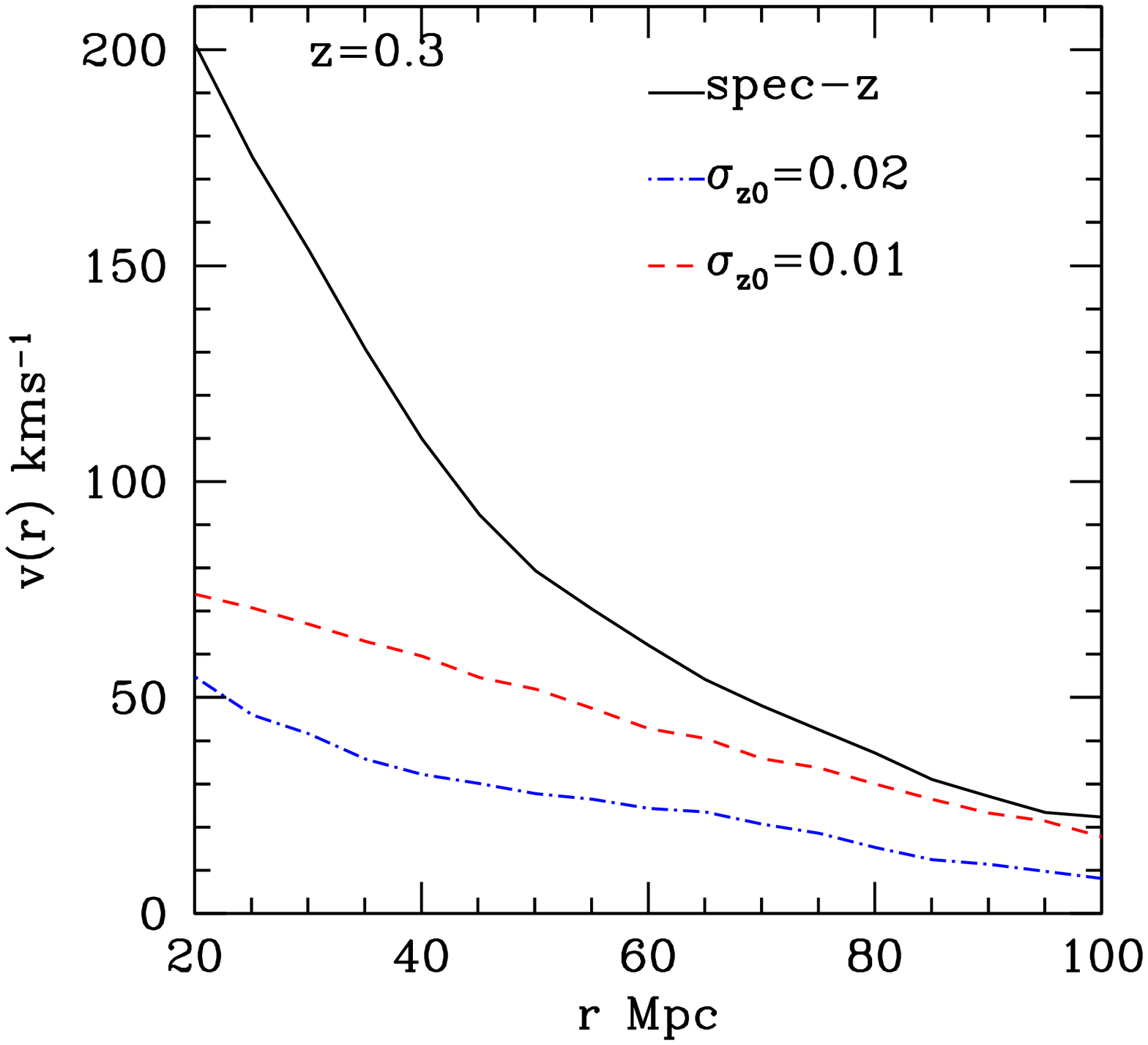}} 
      \resizebox{85mm}{!}{\includegraphics{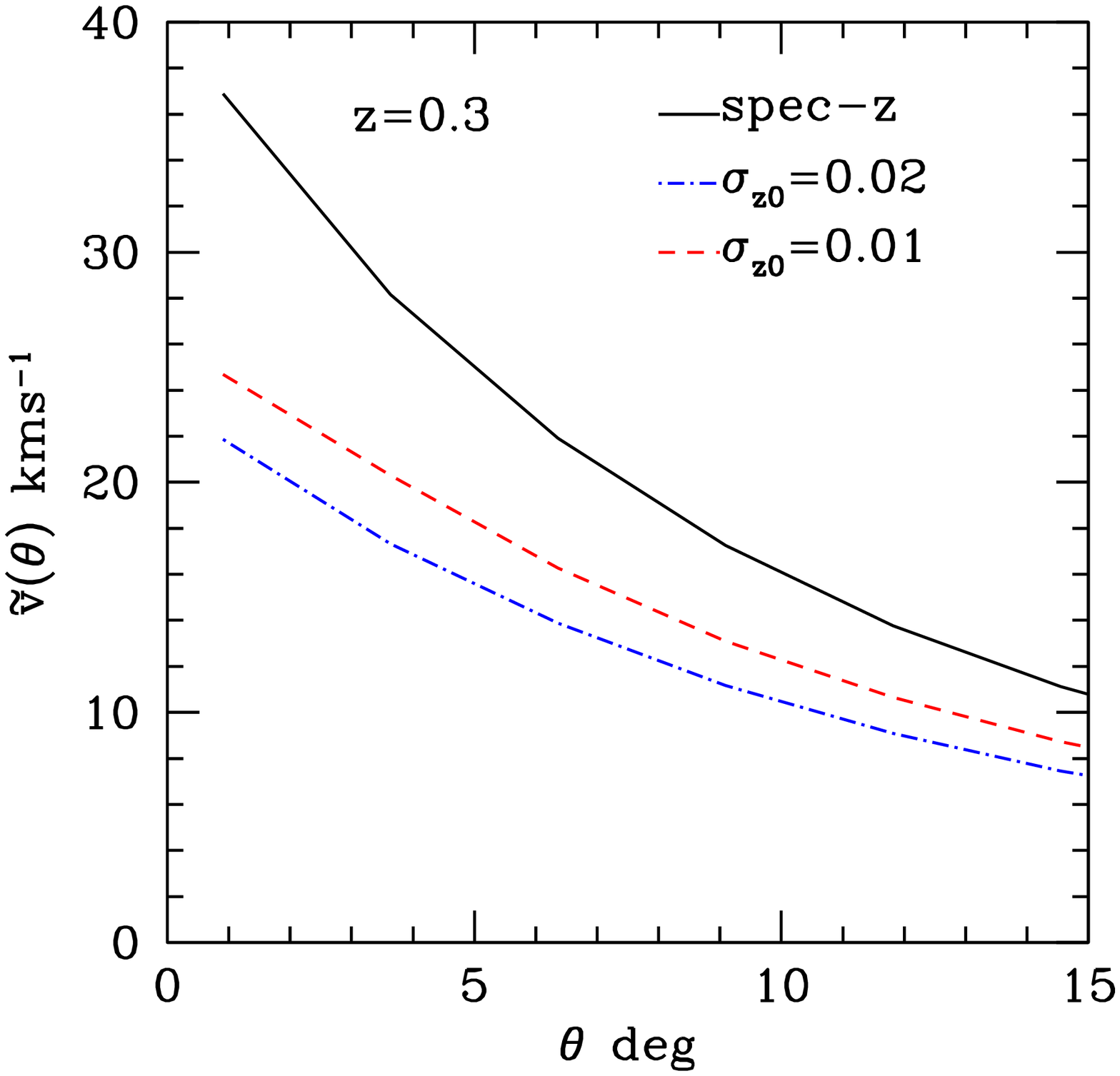}}
     % \resizebox{85mm}{!}{\includegraphics{vivj0.2.eps}}
    \end{tabular}
\caption{
Effect of photo-z errors on the mean pairwise velocity as a function of the 
three-dimensional separation $r$ (left panel) and on the projected mean pairwise velocity as a function 
of angular separation $\theta$ (right panel).   The solid line represents the spectroscopic 
sample where the positions of the SNIa host galaxies are known accurately. The dashed line 
corresponds to a scenario where the dispersion in the photo-z distribution about the true redshift is
given by $\sigma_z=\sigma_{z0}(1+z)$ with $\sigma_{z0}=0.01$, whereas the dot-dashed line represents 
the case when $\sigma_{z0}=0.02$.  Projected statistics vary less with $\sigma_{z0}$, so they
are less sensitive to systematic errors in this quantity.}
    \label{fig: photoz}
  \end{center}
\end{figure*}
%%%%%%%%%%%%%%%%%%%%%%%%%%%%%%%%%%%%

The left panel of Figure~\ref{fig: photoz} shows the effect of photo-z errors on mean
pairwise velocity measurements as a function of three-dimensional separation $r$. 
For a photo-z error $\sigma_{z0}=$ in the range 0.01 to 0.02, the overall amplitude of the mean 
pairwise velocity is suppressed by a factor of 3 to 4 at 
separations $r \le 50\,{\rm Mpc}/h$ compared to the case where all redshifts 
are known perfectly.  
As the separation increases, this suppression becomes less prominent.  This 
is largely because the three-dimensional separations of the SNeIa are uncertain by an 
amount given by the photo-z error, which may be large compared to the three-dimensional 
distances when the separation is small.  For pairs which are farther apart (and often have 
distances dominated by their transverse separations), this smearing effect has much less impact.  
The right panel of Fig.~\ref{fig: photoz} shows the projected mean pairwise velocity as a 
function of angular separation for different assumed photo-z errors.  
Note that because of the integration along the line of sight, 
changing photo-z errors by a factor of two, from $\sigma_{z0}=0.01$ to $\sigma_{z0}=0.02$, 
causes only a 10\% to 15\%  change in the amplitude of the statistic.

%%%%%%%%%%%%%%%%%%%%%%%%%%%%%%%%%%%%%%%%%%%%%%%%%%%%%%%%%%%%%%%%%%%%%%%%%%%%%%%%%%%%%%%%%%%%%%%%%%%
\subsection{Evolution in the Luminosities of Type Ia Supernovae}
\label{subsec:sys_err}

Evolution in intrinsic SNIa properties is one of the most important 
potential sources of systematic error that could 
bias estimates of cosmological parameters from pairwise velocities.  
For instance, the mean intrinsic luminosity of SNeIa could vary 
significantly over time. If this were unaccounted for, the inferred distance moduli 
for supernovae would have a systematic error whose amplitude is a function of redshift. 
Following Ref.~\citep{detf}, we model an error in the distance modulus $\mu$ as
\begin{equation}
\mu= \mu_{\rm true} + \Delta\mu = \mu_{\rm true} + \mu_L  z + \mu_Q z^2, 
\label{eq:musys}
\end{equation}
where $\mu_L$ and $\mu_Q$ are parameters quantifying the linear and quadratic
dependence of the systematic error on redshift.
A systematic error in distance modulus propagates into an error in
the inferred cosmological redshift $z(\mu)$ in Eq.~(\ref{vlos}),
while a systematic error in the photometric redshift directly affects $z_{\rm meas}$. 

Propagating errors through Eqs.~(\ref{mudef}) and (\ref{dLz}) gives
\begin{equation}
\Delta z = \frac{0.46\Delta\mu(1+z)}{f(z)}, 
\label{deltaz}
\end{equation}
where we define the function
\begin{equation}
f(z) \equiv 1 + \frac{(1+z)^2}{d_L(z)H(z)} .
\label{fzdef}
\end{equation}
The resulting error in the line-of-sight velocity for a given supernova is
\begin{equation}
\Delta v_{\rm los} = \frac{0.46c\Delta \mu}{f(z_{\rm meas})}.
\label{deltav}
\end{equation}
Using the measured redshift instead of the cosmological redshift in
this expression gives an error on the order of a few percent at redshifts
of interest.

This systematic shift can be applied directly to the estimator Eq.~(\ref{v_est}) to
evaluate the impact systematic errors have upon a given supernova velocity catalog.
Alternately, we can apply this systematic error to Eq.~(\ref{eq:vij}) to get
an estimate of the size of the resulting shift in the pairwise velocity statistic.
Consider a pair of supernovae with measured redshifts $z_1$ and $z_2$. Each of them has their
three-dimensional velocity systematically shifted in the line-of-sight direction by an amount
$\Delta v_{\rm los}$; the component of this shift along the vector connecting the two
galaxies is $\Delta v_{\rm los} \pi_t/r$ where $\pi_t$ is their separation along the line
of sight and $r$ is the distance between the galaxies. Their pairwise velocity
gets a systematic shift given by
\begin{equation}
\Delta v(r,a) = \frac{c\pi_t}{r}\left(\frac{\Delta z_2}{1+z_2} - \frac{\Delta z_1}{1+z_1}\right)
= \frac{0.46c\pi_t}{r}\left(\frac{\Delta\mu_2}{f(z_2)} - \frac{\Delta\mu_1}{f(z_1)}\right)
\simeq \frac{0.46 H(z_1) \pi_t^2}{f(z_1)r}(\mu_L+2z_1\mu_Q),
\label{deltavij}
\end{equation}
where for the last expression we have used the fact that the difference in the second expression
is dominated by the difference
in distance modulus, rather than the much smaller difference in $f(z)$. In replacing both redshifts by $z_1$ in
this expression, we have assumed that the redshift difference for a given pair is small compared
to unity, which will be the case for any pair separations for which the mean pairwise velocity is significant.
For a given value of $r$ and $a=(1+z_1)^{-1}$,
the only quantity which varies between different pairs 
is the line-of-sight separation term, $\pi_t^2$, whose average over random pairs is $r^2/2$. 
Averaging the final expression in Eq.~(\ref{deltavij}) over all pairs with a given separation 
replace $\pi_t^2/r$ by $r/2$ and gives the systematic error
in the mean pairwise velocity for pairs with comoving separation $r$ and mean redshift $z$ as
\begin{equation}
\Delta v(r,z) = 0.23 r\frac{H(z)(\mu_L +2z\mu_Q)}{f(z)} .
\label{deltavpair}
\end{equation}

For the systematic error in the projected statistic, 
we substitute Eq.~(\ref{deltavpair}) into
Eq.~(\ref{v_projected}), which yields
\begin{equation}
\Delta {\tilde v}(\theta,z) = 0.23\frac{H(z)}{f(z)}(\mu_L + 2\mu_Q z)\int_0^{\pi_{\rm max}}
d\pi_t P(\pi_t | \theta, z) \sqrt{\theta^2 d_C(z)^2 + \pi_t^2} ,
\label{deltatildevpair}
\end{equation}
with $P(\pi_t | \theta, z)$ given by Eq.~(\ref{Ppitheta}). This expression is used in Sec.~\ref{sec:results}
to estimate how small this systematic error must be so that it does not dominate the statistical
errors in mean pairwise velocity measurements of dark energy parameters.

%
%%%%%%%%%%%%%%%%%%%%%%%%%%%%%%%%%%%%%%%%%%%%%%%%%%%%%%%%%%%%%%%%%%%%%%%%%%%%%%%%%%%%%%%%%%%%%%%%%%%%%%%%%
\section{Statistical Errors}
\label{sec: stat_err} 

The line-of-sight velocity for a supernova is inferred by combining a redshift measurement and a distance
estimate obtained from a brightness measurement. Here we assume Gaussian random errors
for both the redshift and brightness measurements, and find the resulting statistical error
in the mean pairwise velocity. We also give an expression for the sample variance 
(sometimes referred to as cosmic variance) error in this quantity,
which results from the fact that its intrinsic value 
in the limited volume we probe may not match the universal mean.

\subsection{Apparent Magnitude and Redshift Errors}
\label{subsec: mea_err}

For a given supernova, we assume normal errors of $\sigma_\mu$ and $\sigma_z$ on the
distance modulus and the measured redshift. Propagating through Eq.~(\ref{vlos})
using Eq.~(\ref{deltaz}) and adding the resulting errors in quadrature gives
\begin{equation}
\delta v_{\rm los}(z)^2 = \frac{0.21 c^2}{f(z)^2}\sigma_\mu^2 + \frac{c^2}{(1+z)^2}\sigma_z^2. 
\label{deltavlos}
\end{equation}
In evaluating the first term, we have assumed that $\sigma_z$ is small compared to $z_{\rm meas}$, which
should be a good approximation for photometric redshifts of SNeIa \citep{pinto05,frieman09}. This allows
us to neglect the effect of errors in $z_{\rm meas}$ on the value of $f(z)$. Note that in actual measurements 
the errors in photometric redshifts may be significantly non-Gaussian, requiring a more 
sophisticated treatment; here we explore Gaussian errors to give an approximate guideline 
for the relevant levels of uncertainty.

Gravitational lensing may increase the dispersion in the measured distance moduli 
of SNeIa beyond that of intrinsic luminosity scatter and random measurement errors.
In the weak lensing limit (convergence $\kappa \ll 1$), 
the dispersion due to lensing is 
\citep{bernardeau_etal97,valageas00,dodelson_vallinotto06,zentner_bhattacharya09}
\begin{equation}
\label{eq:lens}
\sigma_{\rm lens}^2(z) \approx 1.69\Omega_m^2H_0^2\int_0^z dz' \frac{W^2(z',z)}{H(z)} \int dk\,k P(k,z'),
\end{equation}
where $W(z',z)=H_0 d_A(z') d_A(z',z)/d_A(z)$, $d_A(z)$ is the 
angular diameter distance to redshift $z$, and 
$d_A(z',z)$ is the angular diameter distance between redshifts $z'$ and $z$.  The quantity 
$P(k,z)$ is the matter power spectrum; we evaluate it using the numerical fits of \citet{smith_etal03}.  
We thus have a total standard error on the distance modulus for a single supernova composed of three pieces:
\begin{equation}
\sigma_\mu^2 = \sigma_{\rm obs}^2 + \sigma_{\rm SN}^2 + \sigma_{\rm lens}^2 ,
\label{sigmamu_pieces}
\end{equation}
where $\sigma_{\rm obs}$ is the random scatter due to measurement noise and $\sigma_{\rm SN}$
is the intrinsic scatter in supernova intrinsic luminosity. Where not otherwise specified, we take $\sigma_{\rm SN} = 0.1$ independent
of redshift, following recent estimates \cite{detf}, and assume $\sigma_{\rm obs} \ll \sigma_{\rm SN}$,
which should be satisfied for upcoming large surveys like LSST. 

% FOLLOWING 2 PARAGRAPHS REWRITTEN:

To obtain the standard error in the mean pairwise velocity, we begin by assuming that each 
individual line-of-sight velocity has a normally-distributed error with standard deviation $\delta v_{\rm los}$. 
Then for any data bin, applying standard propagation of errors to Eq.~(\ref{v_est}) gives:
\begin{equation}
\delta v^{\rm est} 
%= \frac{\sqrt{2} \delta v_{\rm los} \left(\sum_{\text{pairs}}p^2_{ij}\right)^{1/2}}{\sum_{\text{pairs}}p^2_{ij}}
= \sqrt{2} \delta v_{\rm los} \biggl(\sum_{\text{pairs}}p^2_{ij}\biggr)^{-1/2},
\label{deltav_est1}
\end{equation}
assuming that fractional errors in the $p_{ij}$ are modest; we expect this to hold, 
as these values can be evaluated using redshift distances, 
rather than the comparatively uncertain distance measurements that drive the uncertainty in individual speeds.
For each pair, $p_{ij}^2\simeq \cos^2\varphi$, where $\varphi$ is the angle between the comoving line-of-sight vector and
the vector connecting the comoving supernova positions. This angle will be distributed randomly for each pair; the
expected mean value of $p_{ij}^2$ over a large number of pairs is $0.5$. 
Thus the standard error in the mean pairwise velocity in a particular redshift and separation 
bin is
\begin{equation}
\delta v^{\rm est}(r,a) = 2\frac{\delta v_{\rm los}(a)}{\sqrt{N(r,a)}},
\label{deltavpairwise}
\end{equation}
where $N(r,a)$ is the total number of pairs used to estimate the mean pairwise velocity
in a given redshift bin with mean scale factor $a$ and separation bin with mean separation $r$. 

The standard error on the projected statistic can be expressed as a sum
over pairs in the same way as $v(r,a)$, except the sum is over $N(\theta,z)$ pairs
in a given angular separation bin about $\theta$ instead of a given real-space separation $r$. 
The same calculation applies, except that now the average value of $p_{ij}^2$ for a bin in $\theta$
will not be 0.5. For a given pair, the projector $p_{ij} = \pi_t/r$, where $\pi_t$ is the comoving radial
separation of the pair (as defined in section \ref{sec: theory}). Analogous to Eq.~(\ref{deltav_est1}), 
the error on the projected statistic in a bin can be written as
\begin{equation}
\delta v^{\rm est} 
= \sqrt{2} \delta v_{\rm los} \biggl(\sum_{\text{pairs}}\frac{\pi_t^2}{r^2}\biggr)^{-1/2}.
\label{deltavtilde_est1}
\end{equation}
The sum must be evaluated by integrating over all the pairs in a given angular bin, giving
\begin{equation}
\delta{\tilde v}(\theta,z) = \delta v_{\rm los}(z)\left[\frac{2}{N(\theta,z)}\int_0^{\pi_{\rm max}} d\pi_t
P(\pi_t|\theta, z)\frac{\pi_t^2}{\theta^2 d_C(z)^2 + \pi_t^2}\right]^{-1/2},
\label{deltavpairproj}
\end{equation}
with $P(\pi_t | \theta, z)$ given by Eq.~(\ref{Ppitheta}). 

For a bin in angle covering a range from $\theta_{\rm low}$ to $\theta_{\rm high}$ and
a mean redshift bin with a range from $z_{\rm low}$ to $z_{\rm high}$, we can derive the
number of pairs in this bin from Eq.~(\ref{snrate}). Consider a supernova at redshift $z_1$. 
Any second supernova which lies in the angular bin will be contained in a sky region with
area $2\pi(\cos\theta_{\rm low} - \cos\theta_{\rm high}) \simeq \pi(\theta_{\rm high}^2 - \theta_{\rm low}^2)$,
where the second expression is valid for small angles. The second supernova at redshift $z_2 \geq z_1$
must satisfy $z_{\rm low} \leq (z_1 + z_2)/2 \leq z_{\rm high}$ for the pair to be in the redshift bin,
and $c(z_2 - z_1)/H(z_1) < \pi_{\rm max}$ for the comoving line-of-sight separation to be less
than $\pi_{\rm max}$. These conditions are equivalent to
\begin{equation}
z_{\rm low} - \frac{\pi_{\rm max}H(z_{\rm low})}{2c} < z_1 < z_{\rm high} ~~~{\rm and}
\label{z1range}
\end{equation}
\begin{equation}
2z_{\rm low} - z_1 < z_2 < {\rm min}\left[2z_{\rm high} - z_1, \, z_1 + \frac{\pi_{\rm max} H(z_1)}{c}\right].
\label{z2range}
\end{equation}
Then neglecting the effect of any spatial clustering of supernovae, the total number in the
bin is simply
\begin{equation}
N(\theta_{\rm low},\theta_{\rm high};z_{\rm low},z_{\rm high}) \simeq 
\pi(\theta_{\rm high}^2 - \theta_{\rm low}^2)\int dz_1\frac{d^2n}{dz\,d\Omega}(z_1) \int dz_2 \frac{d^2n}{dz\,d\Omega}(z_2) ,
\label{Nbin}
\end{equation}
where the limits on the $z$ integrals are given in Eqs.~(\ref{z1range}) and (\ref{z2range}); note the
$z_2$ integral must be performed first since its limits depend on $z_1$. The function 
$d^2n/dz d\Omega$ is just Eq.~(\ref{snrate}) normalized
to the total number of supernovae assumed per unit solid angle on the sky.

\subsection{Sample Variance}
\label{subsec:cos_err}

In addition to the measurement errors for individual galaxy velocities, there is an additional uncertainty in comparing 
estimates of the mean pairwise velocity to models, resulting from the fact that we only 
sample a finite volume in which the realized average pairwise velocity may 
differ from the mean taken over the entire Universe.  
Here we give an expression for the covariance between the 
projected mean pairwise velocity measured in different redshift and angular 
separation bins resulting from this effect (generally referred to as sample or cosmic variance).

\begin{widetext}

Consider a mean pairwise velocity statistic binned in pair separation, 
$r$, and scale factor, $a$.
For the three-dimensional mean pairwise velocity, Eq.~(\ref{v12}), the sample
covariance between two bins in separation and scale factor 
$[r,a]_m$ and $[r,a]_n$ for a survey volume $V_\Omega$ can be written as \cite{bk07}
\begin{eqnarray}
C(r_m,r_n;a_m,a_n)&=&\frac{32\pi H(a_m)a_m H(a_n)a_n}{9V_\Omega (1+\xi^{\rm gal}(r_m,a_m))(1+\xi^{\rm gal}(r_n,a_n))}\nonumber\\
&&\qquad\qquad\qquad\qquad
\times\left(\frac{d \ln D_a}{d \ln a}\right)_{a_m}\left(\frac{d \ln D_a}{d \ln a}\right)_{a_n} 
\int dk k^2 |P(k)|^2j_1(kr_m)j_1(kr_n).
\label{C_sample}
\end{eqnarray}
We now integrate along the line-of-sight 
accounting for the photo-z errors and obtain an expression for the sample covariance of the
projected mean pairwise velocity as a function of perpendicular separation,
\begin{eqnarray}
C(\theta_m,\theta_n;a_m,a_n)&=& \int_0^\infty d\pi^{(m)}_{\rm obs}
\int_0^\infty d\pi^{(m)}_{\rm t} P(\pi^{(m)}_{\rm t}|\theta_m,a_m)
P(\pi^{(m)}_{\rm obs}|\pi^{(m)}_{\rm t})\nonumber\\
&&\qquad\qquad\qquad
\times\int_0^\infty d\pi^{(n)}_{\rm obs}
\int_0^\infty d\pi^{(n)}_{\rm t} P(\pi^{(n)}_{\rm t}|\theta_n,a_n)
P(\pi^{(n)}_{\rm obs}|\pi^{(n)}_{\rm t}) C(r_m,r_n;a_m,a_n),
\label{C_proj}
\end{eqnarray}
using Eqs.~(\ref{Ppitheta}) and (\ref{separation_los}). 
\end{widetext}

The total statistical error covariance matrix is the sum of the sample covariance,
Eq.~(\ref{C_proj}), and the statistical error, Eq.~(\ref{deltavpairproj}):
\begin{equation}
C_{\rm total}(\theta_m,\theta_n;a_m,a_n)=C(\theta_m,\theta_n;a_m,a_n)
+\delta_{mn} \delta{\tilde v}^2(\theta_m,a_m) .
 \label{C_vpij_t}
\end{equation}
In the following Section, we use this total covariance matrix to estimate the observability of 
SNeIa peculiar velocities and their utility to cosmology.

%%%%%%%%%%%%%%%%%%%%%%%%%%%%%%%%%%%%%%%%%%%%%%%%%%%%%%%%%%%%%%%%%%%%%%%%%%%%%%%%%%%%%%%%%%%%%%%%%%%%%%%%
\section{Results}
\label{sec:results}

%%%%%%%%%%%%%%%%%%%%%%%%%%%%%%%%%%%%%%%%%%%%%%%%%%%%%%%%%%%%%%%%%%%%%%%%%%%%%%%%%%%%%%%%%%%%%%%%%%%%%
\subsection{The Signal-To-Noise Ratio of Projected Mean Pairwise Velocity Measurements}
\label{subsec:StoN}

\begin{figure*}
  \begin{center}
    \begin{tabular}{c}
      \resizebox{85mm}{!}{\includegraphics{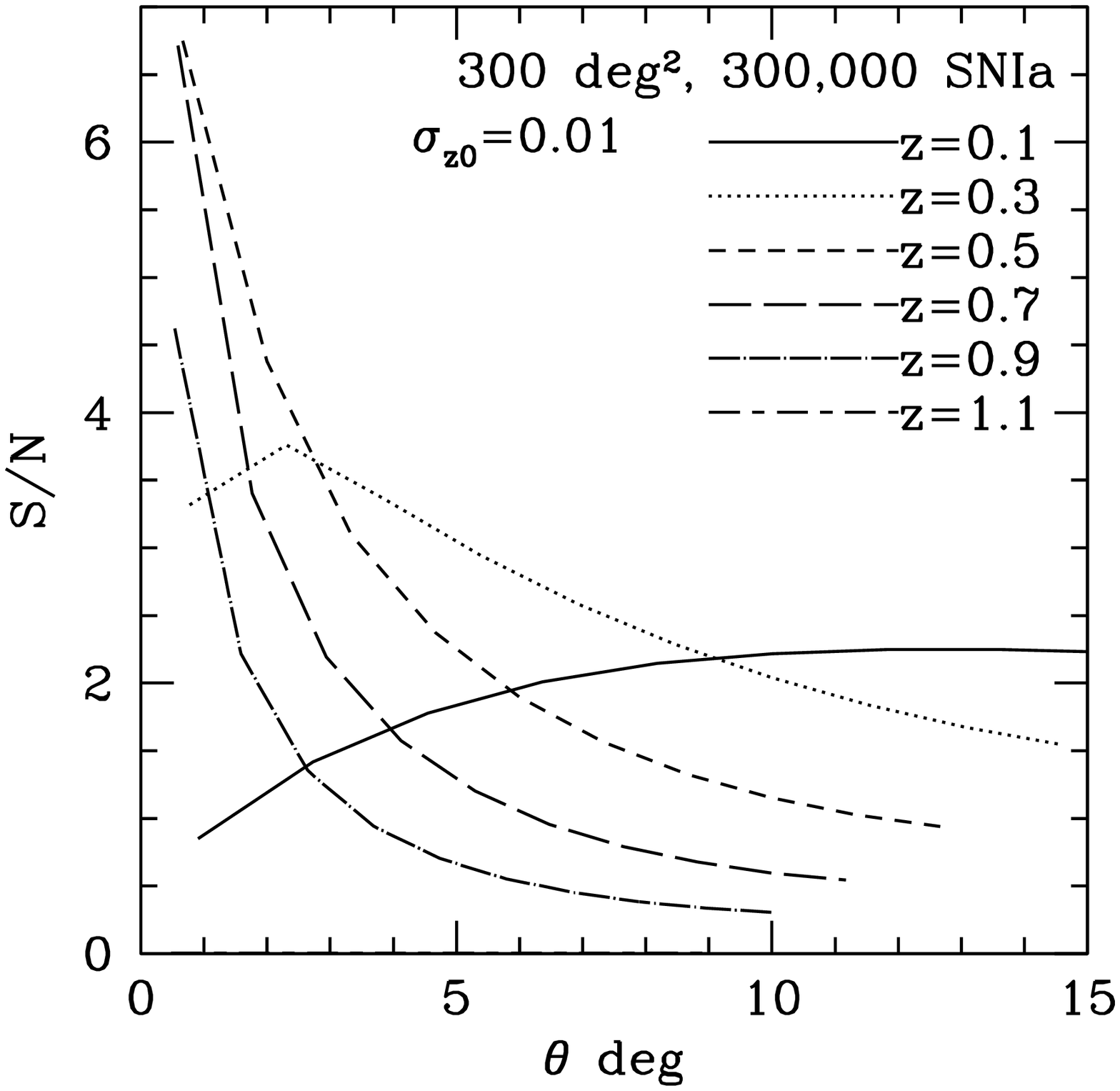}} 
      \resizebox{85mm}{!}{\includegraphics{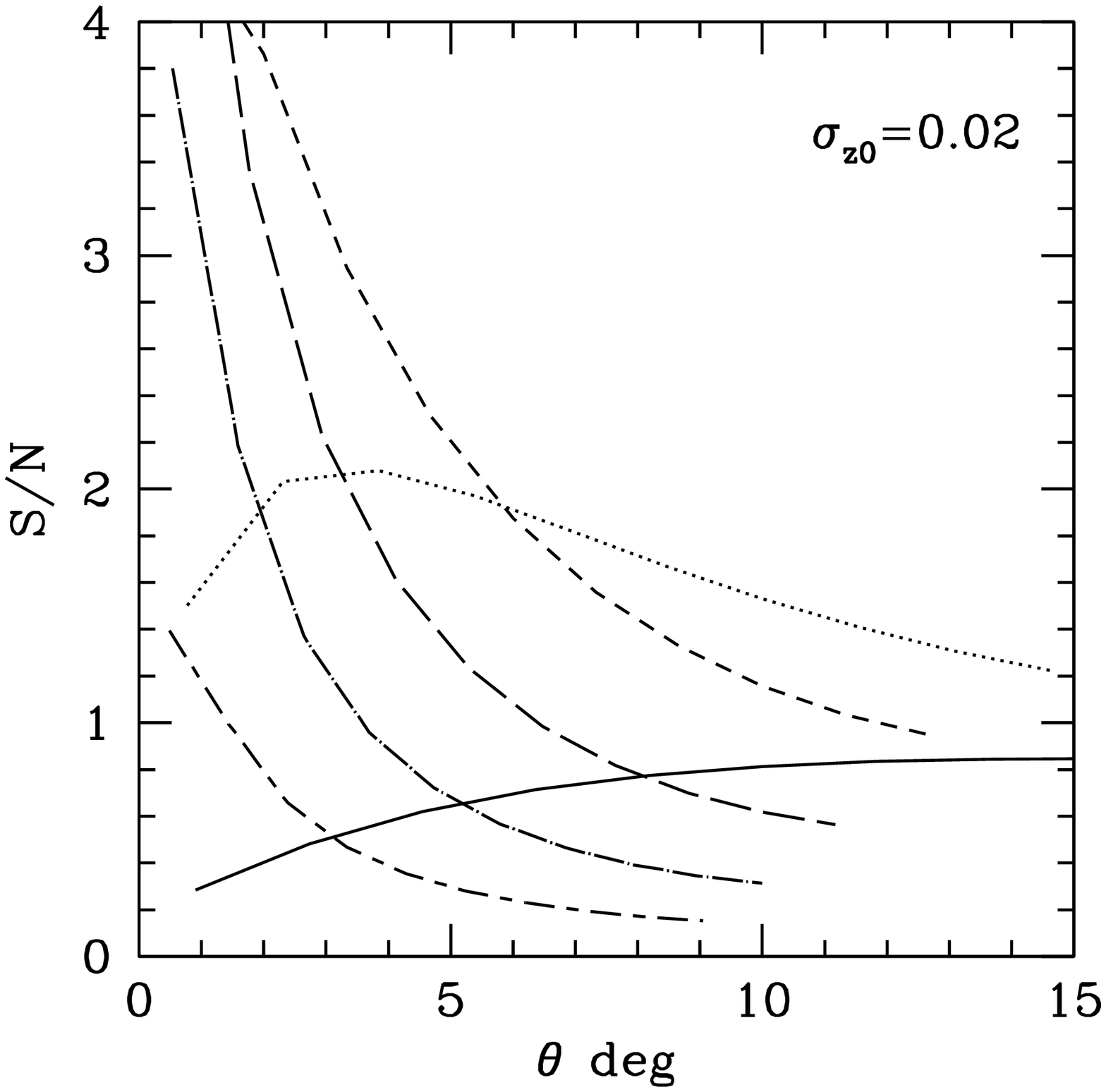}}\\
     \resizebox{85mm}{!}{\includegraphics{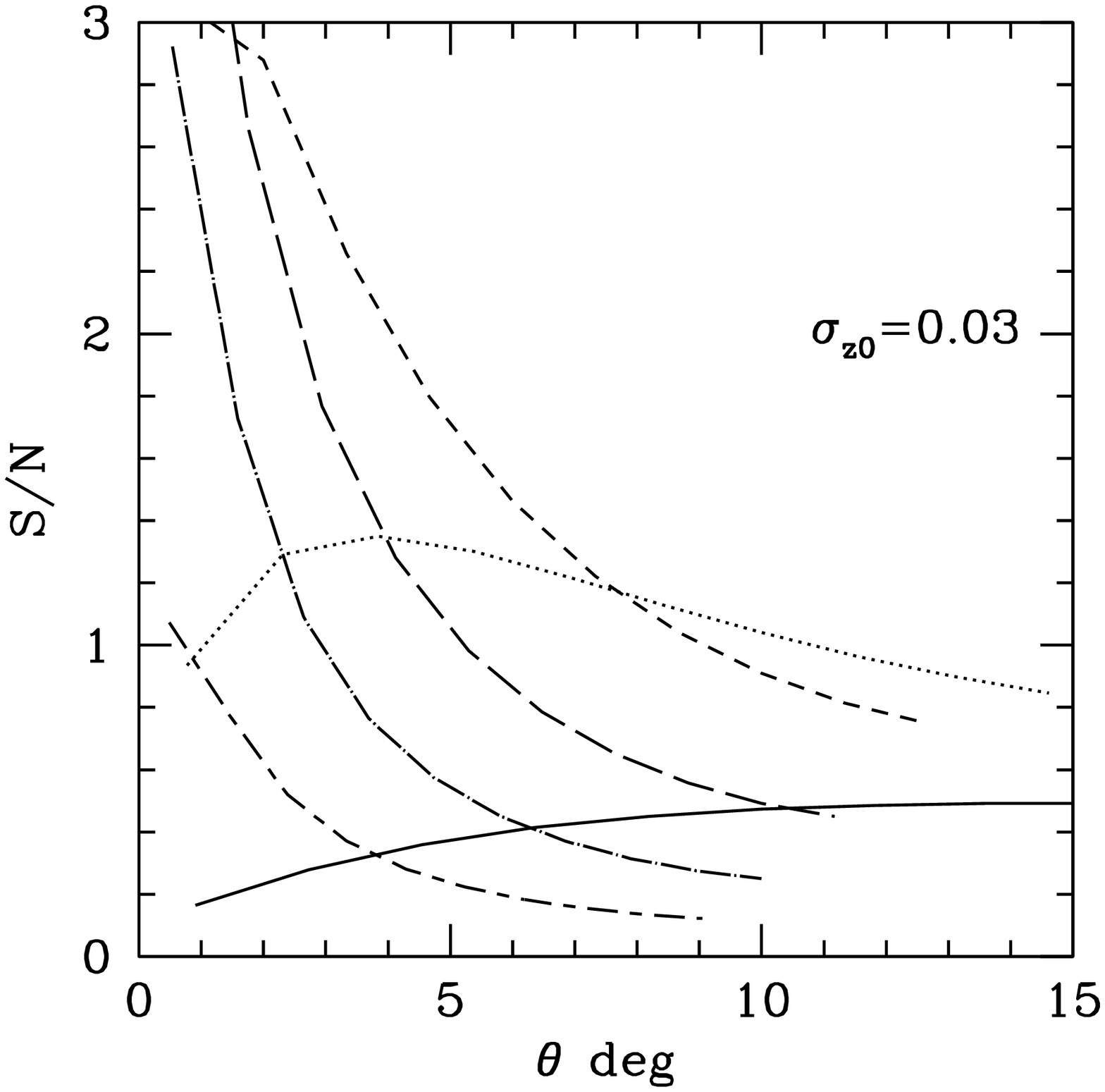}}
     \resizebox{85mm}{!}{\includegraphics{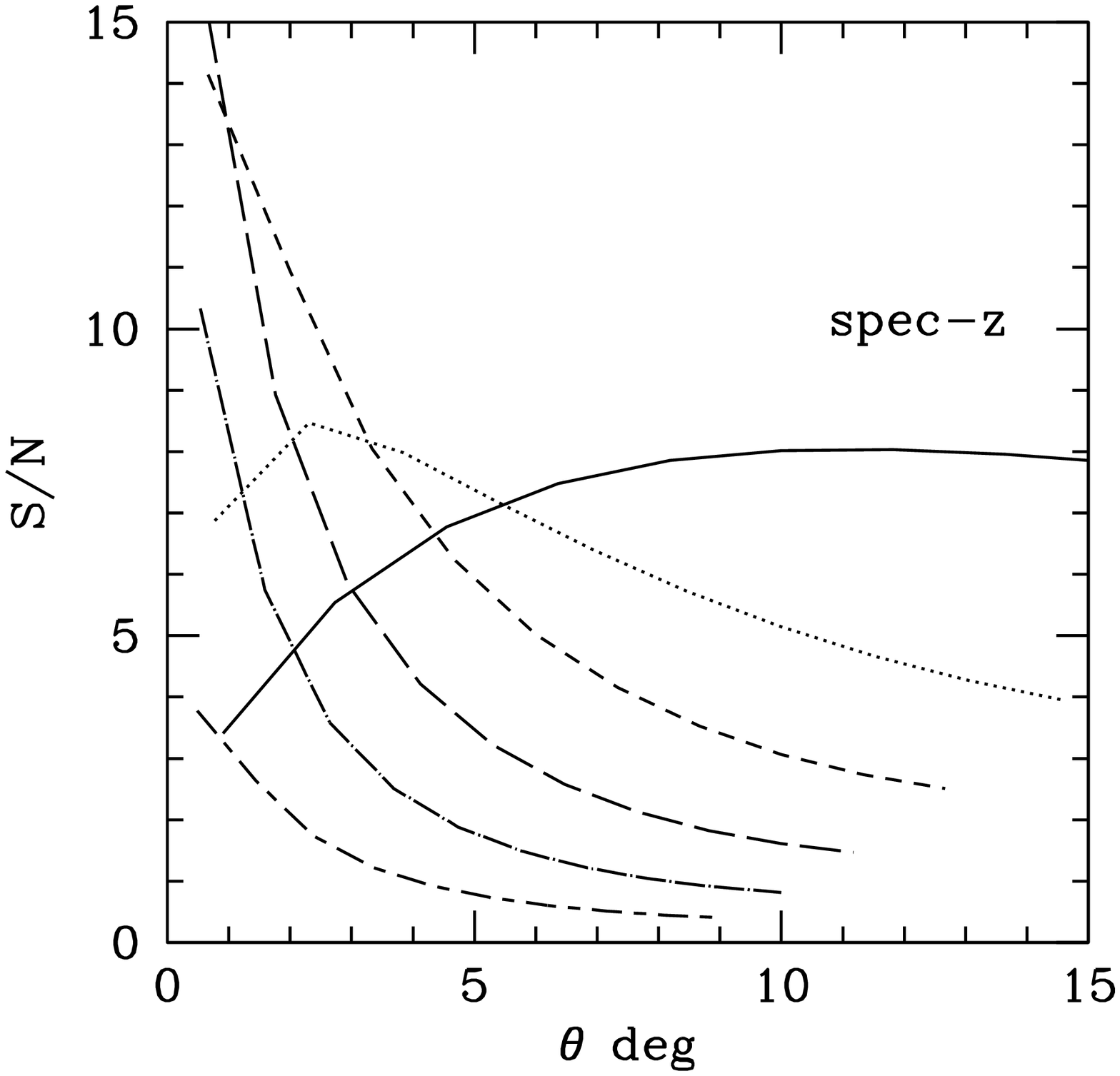}}
    \end{tabular}
    \caption{
The signal-to-noise per angular bin for the projected mean pairwise velocity for different redshift 
bins as a function of angular separation $\theta$, for a catalog with $3\times 10^5$ total
host galaxies over 300 square degrees of sky, and a distance modulus
scatter for each host galaxy of $\sigma_{\rm SN} = 0.1$ plus the scatter due to lensing magnification.  
The pairs are binned by their 
photometric redshifts and the central values of the redshift bins are shown; each 
redshift bin has a width of $\Delta z=0.2$. The maximum angle for each redshift bin 
corresponds to the angle subtended by 100 Mpc at the bin's mean redshift; the range
in angles from 0 to the maximum angle is divided into 10 angular bins. The top panels show the signal-to-noise for  a 
photo-z normal error given by $\sigma_z= \sigma_{z0}(1+z)$, where 
$\sigma_{z0}= 0.01$ (left) and $\sigma_{z0}= 0.02$ (right). The lower left (right) panel 
shows the signal-to-noise for $\sigma_{z0}= 0.03$ (left) and the ``spectroscopic'' limit 
with $\sigma_{z0}=0.001$ (right).
}
\label{fig:StoN}
\end{center}
\end{figure*}

As seen in Fig.~\ref{fig: photoz}, the projected velocity statistic given by Eq.~(\ref{v_projected}) is far less 
sensitive to photometric redshift errors than the non-projected pairwise velocity.  We therefore will 
use this statistic both to estimate the signal-to-noise of pairwise velocity measurements and to determine 
the resulting constraints on cosmological parameters.  The simple pairwise velocity should yield 
comparable or better constraints in the limit of small photometric redshift errors, but the results will be more sensitive to 
$\sigma_z$.

Figure~\ref{fig:StoN} shows the signal-to-noise ratio per angular bin for measurements of 
the projected mean pairwise velocity as a function of angular separation $\theta$,
for our fiducial survey giving $3\times 10^5$ total
host galaxies over 300 square degrees of sky, and a distance modulus
scatter for each host galaxy of $\sigma_{\rm SN} = 0.1$ plus the scatter due to lensing magnification.  
Pairs are binned in 6 redshift bins equally spaced between $z=0$ and $z=1.2$. 
For each redshift bin, 10 bins in angle are used, equally
spaced for angles ranging from $\theta=0$ up to the angle subtended by our maximum
pair separation of 100 Mpc at the mean redshift for the redshift bin.  The maximum angle considered therefore 
decreases as the redshift increases, causing the curves in Fig.~\ref{fig:StoN} to truncate at differing
values of $\theta$.

The  mean pairwise velocity is detectable at a wide range of angular separations and redshifts.  
The top left panel of Figure~\ref{fig:StoN} shows that such a measurement with a 
photometric redshift error of $\sigma(z)=0.01(1+z)$ yields a signal-to-noise ratio between
2 and 9 over a range in angular scales for all but the most extreme redshift bins with $z > 0.8$.  
The redshift distribution of observed SNeIa peaks around $z=0.5$ in our LSST-like model, so 
the closer we get to that redshift range, the more host galaxy pairs we average over 
and the better we can measure velocity statistics.  
Note also that although the number of pairs increases at larger separation, the amplitude 
of the mean pairwise velocity decreases, yielding an overall decrease in the signal-to-noise 
for bins with larger separations.   The top right panel of Fig.~\ref{fig:StoN} shows the signal-to-noise ratio for a photometric redshift error of $\sigma_z=0.02(1+z)$.  After this doubling of the photometric redshift error, the signal-to-noise 
decreases by around 30\%.  Even for $\sigma_z=0.03(1+z)$ (lower left panel of Fig.~\ref{fig:StoN}),
we still reach a signal-to-noise of around 3 for the redshift bins at $z=0.5$ and $z=0.7$.  

The lower right panel shows a best-case scenario, assuming that spectroscopic redshifts are 
obtained for each supernova host galaxy; for simplicity, 
we define a spectroscopic redshift to have $\sigma_{z}=0.001(1+z)$. This redshift error is generally obtainable only from spectroscopy of the hosts (rather than the SNe themselves), primarily because of the large breadth of SNIa spectral features, but also due to the peculiar velocities of SNe with respect to their galaxy's center, which can reach a few hundred km s$^{-1}$.  
Spectroscopic redshifts for large samples of hosts (though likely not all, since many will be fainter than the SNe) would be quite feasible with a 5000--fiber, large field of view multi-object spectrograph like that currently 
proposed for the BigBOSS project \citep{schlegel09}.  If supernova samples cover 300 square degrees, as assumed above, a minimum of 43 BigBOSS pointings would be required to cover this sky region, yielding more than 200,000 redshifts; larger samples can be obtained by revisiting each pointing with different fiber placements.  The proposed BigBOSS survey would use the Kitt Peak 4-meter telescope for only 100 nights per year; such a supernova project would require only a small fraction of the remaining
time available.  In this ``spectroscopic limit,'' the signal-to-noise in measuring the mean pairwise velocity generally improves by around a factor of two compared to the  $\sigma_z=0.01(1+z)$ case.  

%%%%%%%%%%%%%%%%%%%%%%%%%%%%%%%%%%%%%%%%%%%%%%%%%%%%%%%%%%%%%%%%%%%%%%%%%%%%%%%%%%%%%%%%%%%%%%%%%%%%
\subsection{Parameter Space and Formalism}
\label{formalism}

Now we investigate the constraints on dark energy parameters from a SNIa projected mean pairwise
velocity measurement, and assess the complementarity of these constraints to performing
the luminosity distance test based on the same data.
For the sake of simplicity, we perform a Fisher matrix analysis similar to those
in Refs.~\citep{bk07, zentner_bhattacharya09}.  In order to compute constraints 
on $\Omega_\Lambda$, $w_0$ and $w_a$, 
we marginalize over the remainder of the parameter space, consisting of the
parameters $\Delta_\zeta$, $n_S$, and $h$.  We also treat $\sigma_{z0}$,
describing the photometric redshift dispersion, as a parameter since the binned mean
pairwise velocity signal depends on this quantity. 

In addition to the marginalized constraints on $\Omega_{\Lambda}$, $w_0$, and $w_a$, 
we quantify the additional constraining power of pairwise velocities by evaluating 
the quantity $[\sigma(w_p)\sigma(w_a)]^{-1}$ for comparison to  the DETF summary tables \citep{detf}. 
We refer to this as the ``Figure of Merit'' (FoM) for convenience, although in the DETF report
this term refers to a slightly different quantity (the inverse area of the 
$95\%$ confidence limit ellipse in the $w_p-w_a$ plane) which is proportional
to $[\sigma(w_p)\sigma(w_a)]^{-1}$. 
The derived parameter $w_p$ is the equation of state at the 
``pivot'' (i.e. best-constrained) redshift, defined as 
$w_p=w_0+(1-a_p)w_a$ with $a_p= 1+[F^{-1}]_{w_0w_a}/[F^{-1}]_{w_aw_a}$.  

The Fisher matrix for the projected mean pairwise velocity can be 
written as 
\begin{equation}
F_{\alpha\beta}= \sum_{m,n}\frac{\partial {\tilde v}(m)}{\partial p_\alpha}C_{\rm total}^{-1}(mn)\frac{\partial {\tilde v}(n)}{\partial p_\beta} , 
\label{fisher}
\end{equation}
where we have abbreviated the projected mean pairwise velocity in
the nth angular separation and redshift bin as ${\tilde v}(n)$,
$C_{\rm total}(mn)$ is the total covariance matrix between bins $m$ and $n$
given by Eq.~(\ref{C_vpij_t}), and $p_{\alpha}$ indexes
the parameters in the vector ${\bf p}$.  The Fisher 
matrix provides a local estimate of the parameter covariance, 
so the standard error on parameter $p_\alpha$ marginalized over the other parameters is 
$\sigma(p_{\alpha}) = [F^{-1}]_{\alpha \alpha}$ (no summation implied).  

Prior constraints on any of the parameters ${\bf p}$ which are normally distributed are
simple to incorporate. If parameter $p_\alpha$ has a Gaussian prior with standard
error $\sigma_\alpha$, we simply add the diagonal matrix $\rm{diag}(1/\sigma_\alpha^2)$ to the
Fisher matrix $F_{\alpha\beta}$. Priors with non-normal statistical distributions require a
more detailed statistical framework rather than a simple Fisher matrix approximation.

\subsection{Statistical Constraints on Dark Energy Parameters}
\label{constraints}

In computing constraints on the dark energy parameters $\Omega_\Lambda$, $w_0$, and $w_a$,
we first assume a reasonable calibration spectroscopic sample of 1500 SNeIa, comprising
250 supernovae in each redshift bin spread uniformly over the 6 redshift bins spanning $0<z<1.2$.  
The fractional error on the photo-z dispersion, 
$\delta \sigma_{z0}/\sigma_{z0}$ in this case is around $1/\sqrt{500}$, or approximately $5\%$
(assuming Gaussian errors).  We therefore incorporate a Gaussian prior 
on $\sigma_{z0}$ centered on the true value and with 
$\sigma = 0.05 \sigma_{z0}$; however, as we show below in Fig.~\ref{fig: photoz_fisher}, 
the pairwise velocity statistic is relatively 
insensitive to the choice of a prior on $\sigma_{z0}$, so this choice should not significantly 
affect our results, even if the actual error on $\sigma_{z0}$ is much larger.  

We compute the standard errors obtainable on the dark energy parameters using a range of
supernova distance modulus dispersions $\sigma_{\rm SN}$ and photometric redshift dispersions
$\sigma_{z0}$. We consider three possible values of the intrinsic supernova absolute magnitude 
dispersion given by $\sigma_{\rm SN}=0.05$, $\sigma_{\rm SN}=0.1$, 
and $\sigma_{\rm SN}=0.2$.  For each value of $\sigma_{\rm SN}$, we explore four possible values 
of photometric redshift dispersion, $\sigma_{z0}=0.001$ (the ``spectroscopic limit''), 
$\sigma_{z0}=0.01$, $\sigma_{z0}=0.02$, and $\sigma_{z0}=0.03$.  The optimistic but reasonable 
supernova luminosity distance test assumed in the DETF report corresponds to 
$\sigma_{\rm SN}=0.1$ and $\sigma_{\rm z0}=0.01$ so these choices constitute a sensible 
baseline for comparison to other techniques.

The strength of the dark energy constraints obtained is relatively sensitive to the amount of
prior information assumed. First, we can make the same assumptions used 
by the Dark Energy Task Force \citep{detf}.
They assume constraints on all parameters (including covariances) at the 
level expected for measurements of the microwave background power spectrum 
by the Planck satellite. For this, we employ the Planck Fisher Matrix 
provided by the DETF. In addition, DETF assume a 11\% Gaussian 
prior on the value of $h$ \cite{hst}. Note that a spatially flat universe 
is {\it not} assumed. We also assume no systematic error on either redshift or
distance modulus measurements; limits on these systematics required to attain the
statistical error levels presented here are discussed below.  
The results are given in Table~\ref{tab:vijconstraint_detf}. 

For the nominal DETF survey case, mean pairwise velocities give a standard error 
on $w_0$ of $\sigma(w_0)=0.45$ and a standard error on $w_a$ of $\sigma(w_a)=0.98$. 
This constraint on $w_0$ is comparable to the DETF Stage-IV constraints from ground-based 
optical baryon acoustic oscillations
or galaxy cluster counts, while not as good as those from Stage-IV supernova luminosity distances. For
$w_a$, mean pairwise velocity constraints are significantly better than the optical survey-based BAO projection; slightly weaker
than the pessimistic BAO projections for space-based or radio observations and for the optimistic galaxy cluster projection; and halfway between the optimistic and pessimistic DETF supernova luminosity distance projection. However, all of these methods trail the Stage IV weak lensing projections in constraining power.

Among the dark energy probes resulting from a large ground-based optical survey like LSST, mean pairwise
velocities compare well with both the baryon acoustic 
oscillation and the supernova luminosity distance probes \cite{zentner_bhattacharya09}. 
To quantify this, we consider the improvement in dark energy parameters obtained by adding
the mean pairwise velocity probe to the supernova luminosity distance probe resulting from the same sample. 
The mean pairwise velocity can be measured using the supernova data from
a large survey telescope with little additional cost compared to simply constraining
dark energy using the resulting supernova Hubble diagram. 

\begin{figure*}
  \begin{center}
    \begin{tabular}{cc}
      \resizebox{85mm}{!}{\includegraphics{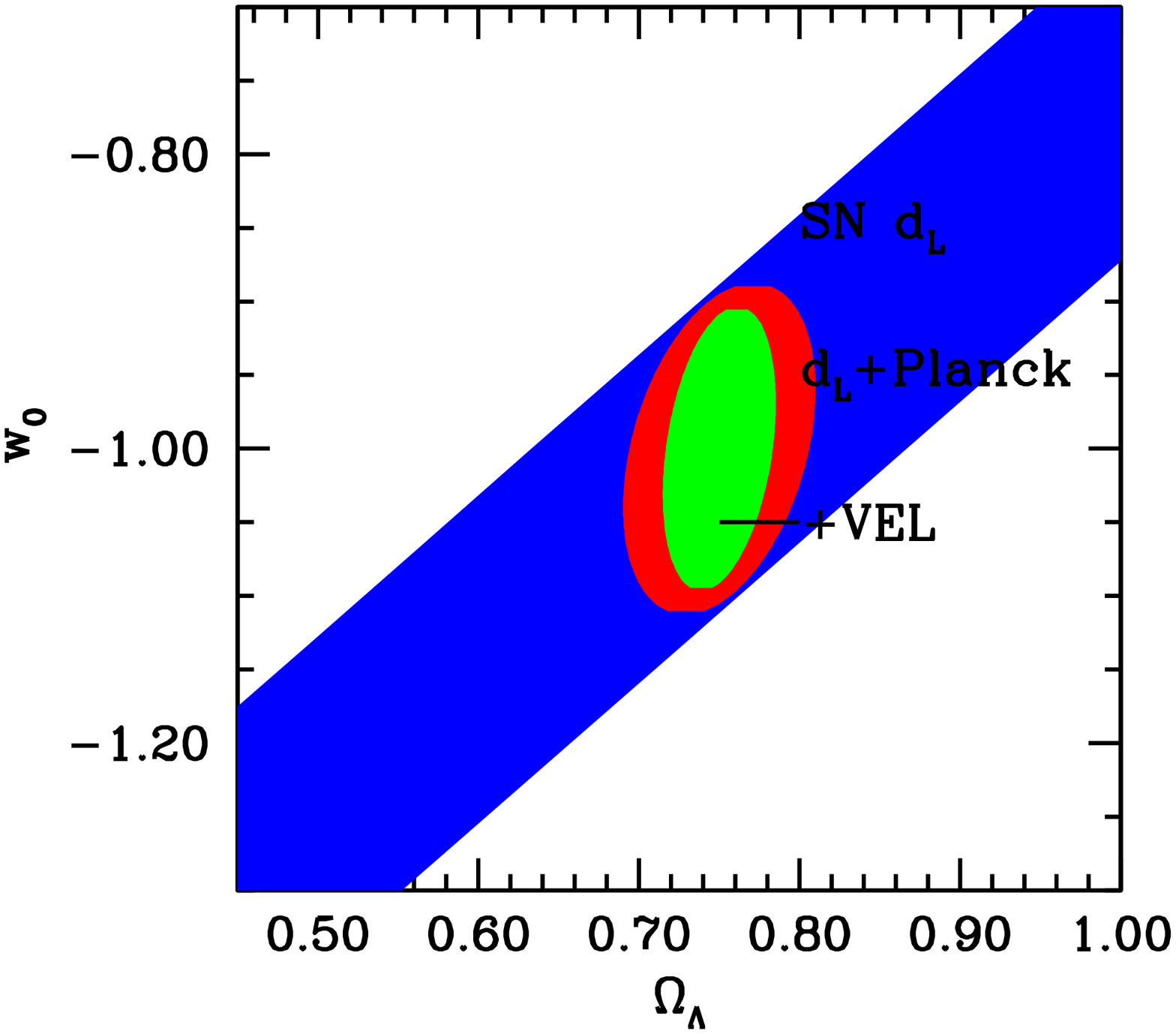}} 
      \resizebox{85mm}{!}{\includegraphics{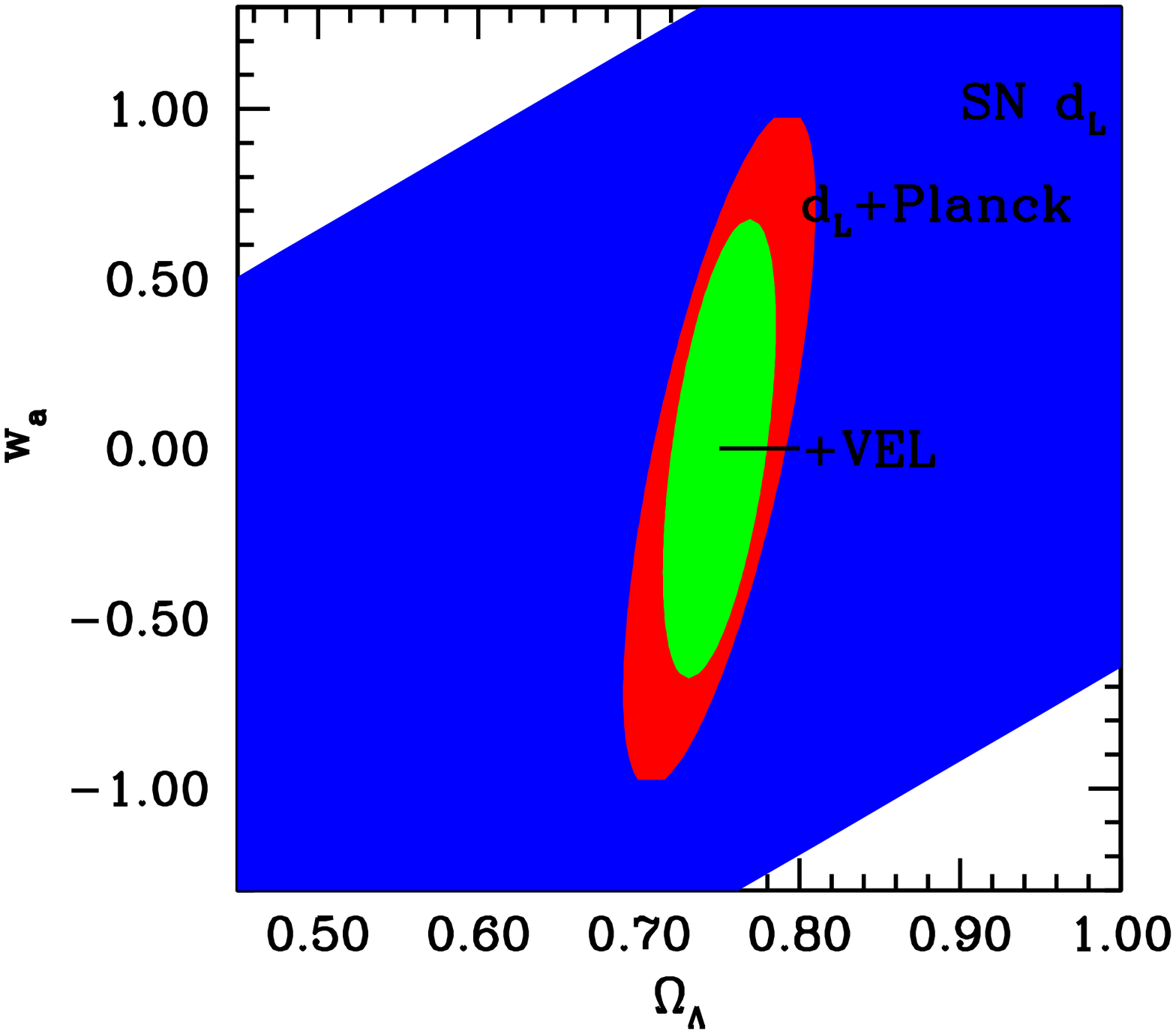}}
     \end{tabular}
    \caption{
Joint dark energy constraints obtainable from an LSST-like future 
supernova survey, combining constraints from the supernovae luminosity 
distance test, priors on Hubble parameter \citep{hst} and constraints from Planck with those obtainable from 
supernova velocity statistics. A flat universe is not assumed. 
A fiducial value for the photometric redshift error of $\sigma_{z0}=0.01$ 
is assumed, with a Gaussian prior on $\sigma_{z0}$ of 5\%. 
The Fisher matrices for the Planck priors and the SNIa 
luminosity distance priors are obtained from the DETF report \citep{detf}. 
The luminosity distance Fisher matrix represents the DETF LSST supernovae 
optimistic (LST-o) survey. The blue (dark) shaded region shows the constraint 
obtainable from the supernova luminosity distance test only. The red (grey) 
region shows the constraint when priors from a Planck survey are combined 
with the distance test. The green (innermost, light shaded) region shows the joint 
constraint combining the distance test, a Planck prior and mean peculiar 
velocity measurements for the SNIa host galaxies.
}
\label{fig: planck_dl_vel}
\end{center}
\end{figure*}

Figure~\ref{fig: planck_dl_vel} shows joint constraints on the dark energy parameters 
combining projected peculiar velocity measurements and the SNIa luminosity distance test, using
the same priors as Table~\ref{tab:vijconstraint_detf}. The left panel shows the 
$1\sigma$ constraint in the $w_0-\Omega_\Lambda$ plane and the right panel 
shows the constraint in the $w_a-\Omega_\Lambda$ plane, after marginalizing over 
the remainder of parameter space.  Incorporating peculiar velocity information significantly 
reduces the size of the ellipses in the dark energy parameter space: the marginalized constraint
on $\Omega_\Lambda$ improves by a factor of 1.7, on $w_0$ by a factor of 1.2, and on
$w_a$ by a factor of 1.5, giving an overall improvement in the Figure of Merit by a factor of 1.8.
(As a point of comparison, corresponding constraints with no priors from other measurements
are included in Fig.~\ref{fig: photoz_fisher}.)
Note that unlike the case of peculiar velocity measurements, the constraints derived from the SNIa 
luminosity distance are sensitive to the error in mean redshift of a bin and hence the 
cosmological constraints derivable from the SNIa luminosity distance depend much more 
on the amount of prior knowledge of the photo-z distribution \citep{huterer04,zentner_bhattacharya09}, 
as well as being much more sensitive to intrinsic SNIa luminosity evolution. 

%%%%%%%%%%%%%%%%%%%%%%%%%%%%%%%%%%%%%%%%%%%%%%%%%%%

%%%%%%%%%%%%%%%%%%%%%%%%%%%%%%%%%%%%%%%%%%%%%%%%%%%%%%%%%%%%%%%%%%%%%%%%%%%%%%%%%%%%%%%%%%%%%%

We have also considered statistical dark energy constraints from a more constraining, but still realistic, set of priors.
In particular, a measurement of the Hubble parameter based on an improved, NGC 4258-calibrated distance ladder
with an estimated overall error of 5\% has recently been reported \cite{rie09}. Furthermore, requiring
the dark energy probe itself in combination with microwave background data to determine the
geometry of the universe is likely overly restrictive. Measurements of the baryon acoustic
oscillation scale from the Sloan Digital Sky Survey Data Release 7, combined with WMAP 5-year
data, give a constraint on the curvature parameter of $\Omega_k = -0.013\pm 0.007$, even
for a very general cosmological model which allows both a nonflat universe and a value of $w_0$ different from -1 \cite{per10}. 
Additionally, since a flat universe is an unstable fixed point for standard
cosmological evolution, we have an overwhelming theoretical prejudice for $\Omega_k=0$ to
high precision. Therefore, a prior assumption of a flat universe is both reasonable and
strongly suggested by data. 

Table~\ref{tab:vijconstraint} gives the standard errors on
the dark energy parameters for a flat universe, with gaussian priors for $h$ (5\%),
$\Delta_\zeta$ (5\%), and $n_S$ (1\%), the latter two being current limits from
WMAP 7-year data \cite{kom10}. The assumption of a flat universe and a tighter prior
on $h$ lead to much stronger dark energy constraints than do DETF priors. 
With these priors, a DETF-assumed supernova sample with $\sigma_{\rm SN}=0.1$ 
and $\sigma_{z0}=0.01$ gives a measurement of $\Omega_\Lambda$, $w_0$ and $w_a$ 
with standard errors of 0.024, 0.27, and 0.41, respectively, using the mean pairwise velocity alone. For comparison, the
constraints on $w_0$ for Stage IV experiments computed in the DETF report (but for the original
set of priors) are worse for clusters, comparable for baryon acoustic oscillations, and better for
the supernova Hubble diagram. Our constraint on $w_a$, on the other hand, is better than for
any of the Stage IV experiments aside from the optimistic weak lensing scenarios. 

Of course, the constraining power of other probes will also increase 
with the more restrictive set of priors we assume in Table~\ref{tab:vijconstraint}.  This makes 
a direct comparison with these other methods beyond the scope of this paper.  Our primary 
point is that under optimistic, but reasonable, assumptions, SNIa 
peculiar velocities can be useful by themselves and at the very least 
can serve as a valuable complementary probe and cross-check for 
systematic errors, while requiring little additional investment.  
However, note that Table~\ref{tab:vijconstraint} shows that broader 
photo-z distributions and/or larger intrinsic SNIa dispersions can
quickly diminish the returns on SNIa peculiar velocities.

This calculation also suggests the potential constraining power of pairwise velocity statistics
from future survey observations. If
a large photometric supernova survey were combined with follow-up
spectroscopic redshifts for supernova host galaxies, the
standard error in the redshift could be reduced by a factor of 10 to $\sigma_{z0}=0.001$,
corresponding to the first column of Table~\ref{tab:vijconstraint}. In this case,
the error on $w_0$ shrinks to $\sigma(w_0) \simeq 0.10$ and the 
error on $w_a$ is nearly $\sigma(w_a) \simeq 0.16$.
Understanding Type-Ia supernovae
well enough to push $\sigma_{\rm SN}$ down by a factor of 2 to $\sigma_{\rm SN}=0.05$ would 
reducethe error on $w_a$ by another factor of two, to 0.08. Few other proposed probes have
comparable potential to constrain $w_a$.

%%%%%%%%%%%%%%%%%%%%%%%%%%%%%%%%%%%%%%
\begin{table}
\begin{center}
\begin{tabular}{| c | ccc | ccc | ccc | ccc |}
\hline
\phantom{a} &
\multicolumn{3}{| c |}{$\sigma_{z0}=0.001$} &
\multicolumn{3}{| c |}{$\sigma_{z0}=0.01$} &
\multicolumn{3}{| c |}{$\sigma_{z0}=0.02$} &
\multicolumn{3}{| c |}{$\sigma_{z0}=0.03$}\\
\hline
$\sigma_{\rm SN}$ & \ \ $\sigma(\Omega_\Lambda)$ & \ \ $\sigma(w_0)$ & \ \ $\sigma(w_a)$ & \ \ 
$\sigma(\Omega_\Lambda)$ & \ \ $\sigma(w_0)$ & \ \ $\sigma(w_a)$ & \ \ $\sigma(\Omega_\Lambda)$ 
& \ \ $\sigma(w_0)$ & \ \ $\sigma(w_a)$ & \ \ $\sigma(\Omega_\Lambda)$ & \ \ $\sigma(w_0)$ & \ \ $\sigma(w_a)$\\
\hline
0.05 &  \ \ 0.016 & \ \ 0.22 & \ \ 0.36 & \ \ 0.032 & \ \ 0.28 & \ \ 0.48 & \ \ 0.05 & \ \ 0.43 & \ \ 0.89 &  \ \ 0.059 & \ \ 0.78 & \ \ 1.7 \\
0.1  & \ \ 0.039 & \ \ 0.30 & \ \ 0.55 & \ \ 0.056 & \ \ 0.45 & \ \ 0.98 & \ \ 0.094 & \ \ 0.62 & \ \ 2.52 & \ \ 0.13 & \ \ 1.46 & \ \ 2.81\\
0.2  & \ \ 0.051 & \ \ 0.42 & \ \ 0.72 & \ \ 0.074 & \ \ 0.59 & \ \ 1.84 &  \ \ 0.18 & \ \ 1.36 & \ \ 4.71 & \ \ 0.24 & \ \ 2.87 & \ \ 6.14\\
\hline
\end{tabular}
\end{center}
\caption{
Dark energy parameter constraints derived from mean pairwise velocity statistics. 
Photometric redshifts are assumed to be normally distributed about the true $z$, 
with $\sigma_z=\sigma_{z0}(1+z)$. We show results for $\sigma_{z0}=0.001$, 
0.01, 0.02, and 0.03, and three different values for the uncertainty in supernova distance moduli, 
$\sigma_{\rm SN}=0.05$, 0.1 and 0.2. 
The fiducial values of the dark energy parameters are $\Omega_\Lambda=0.75$, 
$w_0= -1$ and $w_a= 0$. We assume zero systematic errors related to SNIa evolution, 
i.e.\ $\mu_L=\mu_Q=0$. We assume the same priors used in the DETF report: Planck 
satellite priors from its projected measurement of the microwave background power spectrum 
(using the Fisher matrix supplied by the DETF) and 
a Gaussian prior on $h$ with a standard error of 11\%. This table does not assume 
a flat spatial geometry for the universe.
}
\label{tab:vijconstraint_detf}
\end{table}
%%%%%%%%%%%%%%%%%%%%%%%%%%%%%%%%%%%%

%%%%%%%%%%%%%%%%%%%%%%%%%%%%%%%%%%%%%%
\begin{table}
\begin{center}
\begin{tabular}{| c | ccc | ccc | ccc | ccc |}
\hline
\phantom{a} &
\multicolumn{3}{| c |}{$\sigma_{z0}=0.001$} &
\multicolumn{3}{| c |}{$\sigma_{z0}=0.01$} &
\multicolumn{3}{| c |}{$\sigma_{z0}=0.02$} &
\multicolumn{3}{| c |}{$\sigma_{z0}=0.03$}\\
\hline
$\sigma_{\rm SN}$ & \ \ $\sigma(\Omega_\Lambda)$ & \ \ $\sigma(w_0)$ & \ \ $\sigma(w_a)$ & \ \ $\sigma(\Omega_\Lambda)$ & \ \ $\sigma(w_0)$ 
& \ \ $\sigma(w_a)$ & \ \ $\sigma(\Omega_\Lambda)$ & \ \ $\sigma(w_0)$ & \ \ $\sigma(w_a)$ & \ \ $\sigma(\Omega_\Lambda)$ 
& \ \ $\sigma(w_0)$ & \ \ $\sigma(w_a)$\\
\hline
0.05 &  \ \ 0.004 & \ \ 0.047 & \ \ 0.08 & \ \ 0.012 & \ \ 0.16 & \ \ 0.23 & \ \ 0.024 & \ \ 0.34 & \ \ 0.49 &  \ \ 0.049 & \ \ 0.65 & \ \ 1.03 \\
0.1  & \ \ 0.009 & \ \ 0.10 & \ \ 0.16 & \ \ 0.024 & \ \ 0.27 & \ \ 0.41 & \ \ 0.046 & \ \ 0.62 & \ \ 0.94 & \ \ 0.1 & \ \ 1.42 & \ \ 1.96\\
0.2  & \ \ 0.022 & \ \ 0.28 & \ \ 0.41 & \ \ 0.061 & \ \ 0.63 & \ \ 0.89 &  \ \ 0.1 & \ \ 1.48 & \ \ 1.93 & \ \ 0.2 & \ \ 2.86 & \ \ 4.18\\
\hline
\end{tabular}
\end{center}
\caption{
Same as Table~\ref{tab:vijconstraint_detf}, but we assume $\Omega_k = 0$ and instead of Planck priors we assume 
Gaussian priors with standard errors of 5\% on $h$ and $\Delta_\zeta$ and 1\% on $n_S$ (comparable to errors from current measurements). 
}
\label{tab:vijconstraint}
\end{table}
%%%%%%%%%%%%%%%%%%%%%%%%%%%%%%%%%%%%

\subsection{Systematic Error in Distance Modulus}

%%%%%%%%%%%%%%%%%%%%%%%%%%%%%%%%%%%%
\begin{table}
\begin{center}
\begin{tabular}{| c | ccc |}
\hline
$\mu_L=\mu_Q$ & \ \ $\Omega_\Lambda$ & \ \ $w_0$ & \ \ $w_a$ \\ 
\hline
0.01/$\sqrt{2}$ & \ \ 10.2\% & \ \ 4.1\% & \ \ 20.2\%\\
0.03/$\sqrt{2}$ & \ \ 20.3\% & \ \ 10.0\% & \ \ 41.0\%\\
0.05/$\sqrt{2}$ & \ \ 40.9\% & \ \ 37.5\% & \ \ 80.1\%\\
\hline
\end{tabular}
\end{center}
\label{tab:bias}
\caption{
The ratio of the parameter bias due to systematic error 
to the statistical uncertainties on these parameters. A photometric redshift 
distribution with dispersion $\sigma_z=0.01(1+z)$ is assumed.  
We assume $\mu_L=\mu_Q$ 
to compute the systematic bias. Then we set  $\mu_L=\mu_Q=0$ and compute the statistical uncertainty 
and report the ratio of systematic bias to statistical errors $\Delta p/\sigma_{p}$, where $p=\Omega_\Lambda$, 
$w_0$, or $w_a$ .  We assume $\Omega_k=0$ and assume Gaussian priors with
standard error of 5\% on $h$ and $\Delta_\zeta$ and 1\% on $n_s$. }
\end{table}
%%%%%%%%%%%%%%%%%%%%%%%%%%%

The potential statistical sensitivity of any dark energy probe can only be realized if systematic errors
can be controlled to a level where their effect on cosmological parameters is small compared
to the statistical errors. For the supernova data set considered here, systematic errors
may effect both observables: the distance modulus and the photometric redshift. This section considers 
distance modulus systematics, while the following section analyzes the effect of redshift errors. 

Section~\ref{subsec:sys_err} gives a simple phenomenological model for the effect of SNIa evolution
with redshift, in terms of the parameters $\mu_L$ and $\mu_Q$. 
The resulting systematic error on cosmological parameters induced by 
this systematic error can be estimated using a Fisher matrix 
approach. The bias in parameter $p_\alpha$ can be written as 
\begin{equation}
\delta p_{\alpha}= \sum_{\beta}[F^{-1}]_{\alpha\beta}\sum_{m,n}\Delta{\tilde v}(m)
C_{\rm total}^{-1}(mn)\frac{\partial {\tilde v}(n)}{\partial p_\beta}
\label{fisherbias}
\end{equation}
where $\Delta {\tilde v}$, obtained by substituting Eq.~(\ref{deltavpair}) for $v(r,a)$ in Eq.~(\ref{v_projected}), 
is the systematic shift in the observable $\tilde v$ due to the systematic error characterized
by nonzero values of $\mu_L$ and $\mu_Q$. 

We calculate the bias in each parameter due to SNIa evolution assuming a photometric redshift distribution 
with spread $\sigma_z=0.01(1+z)$ and the evolution model given by Eq.~(\ref{eq:musys}).  
We can then compare the systematic bias with the statistical errors on dark energy parameters
assuming $\mu_L=\mu_Q=0$, as computed in Table~\ref{tab:vijconstraint}.  
The ratios of the bias of the dark energy parameters to their statistical errors are reported in 
Table~III for several representative choices of $\mu_L$ and $\mu_Q$.  
For reference, DETF took 
evolution in SNIa luminosity with $\mu_L=\mu_Q=0.01/\sqrt{2}$ as their optimistic  scenario.  
We find that the maximum bias incurred in $\Omega_{\Lambda}$ and $w_0$ is less 
than 40\% as large as the statistical error on these parameters as long as 
$\mu_L=\mu_Q\le 0.05/\sqrt{2}$ (five times larger than the DETF optimistic systematic error). 
For $w_a$, the systematic bias is 40\% of
the statistical error for $\mu_L=\mu_Q\le 0.03/\sqrt{2}$, and increases to 80\% of the
statistical error for $\mu_L=\mu_Q= 0.05/\sqrt{2}$.  
If the actual unrecognized evolution of SNIa luminosity is similar to that 
assumed in the DETF report, the resulting systematic bias in dark energy parameters
should be insignificant compared to the statistical error. Note that these comparisons 
conservatively use
the statistical error incorporating our more restrictive prior than in the DETF report.
The larger statistical errors with the DETF priors admit substantially larger systematic errors.

\subsection{Systematic Errors in Photometric Redshifts}

We have also tested how a possible bias $\Delta z_p$ in the photo-z distribution
might impact  the dark energy constraints obtainable from pairwise velocity statistics.  
If $\Delta z_p$ is not a strong function of redshift (i.e., it does not vary
considerably within one of our redshift bins with width $\delta z \simeq 0.2$), 
then the bias affects both galaxies in each pair in approximately the same manner.  
The mean pairwise velocity relies on the difference between the two velocities so 
nearly all of the effects of a photo-z bias tend to cancel.  The residual is a 
small misestimation of the location of the redshift bin, which translates into a 
small error in cosmological parameters. For example, assuming a bias in photometric redshifts of 
$\Delta z_p \approx 0.002(1+z)$  degrades the constraints 
on cosmological parameters by less than 2\% of the statistical errors.  
This stands in stark contrast to the strong dependence of the luminosity distance test on 
photometric redshift biases (e.g., \citep{huterer04,zentner_bhattacharya09}) and the similar 
sensitivity of probes such as weak gravitational lensing to biased photometric 
redshifts (e.g., \citep{hearin_etal10}).  

The signal we measure, the redshift-binned projected mean pairwise velocity
Eq.~(\ref{v12perp}), depends on the scatter in photometric redshifts so we also must estimate
the systematic error due to uncertainty in the photometric redshift dispersion. We assume that
the distribution of the difference between photo-z's and spectroscopic redshifts is
a standard normal distribution; in reality this distribution is likely more complex. The
results here are a simple effective model for the distribution of photometric redshifts. 

\begin{figure*}
  \begin{center}
    \begin{tabular}{cc}
      \resizebox{85mm}{!}{\includegraphics{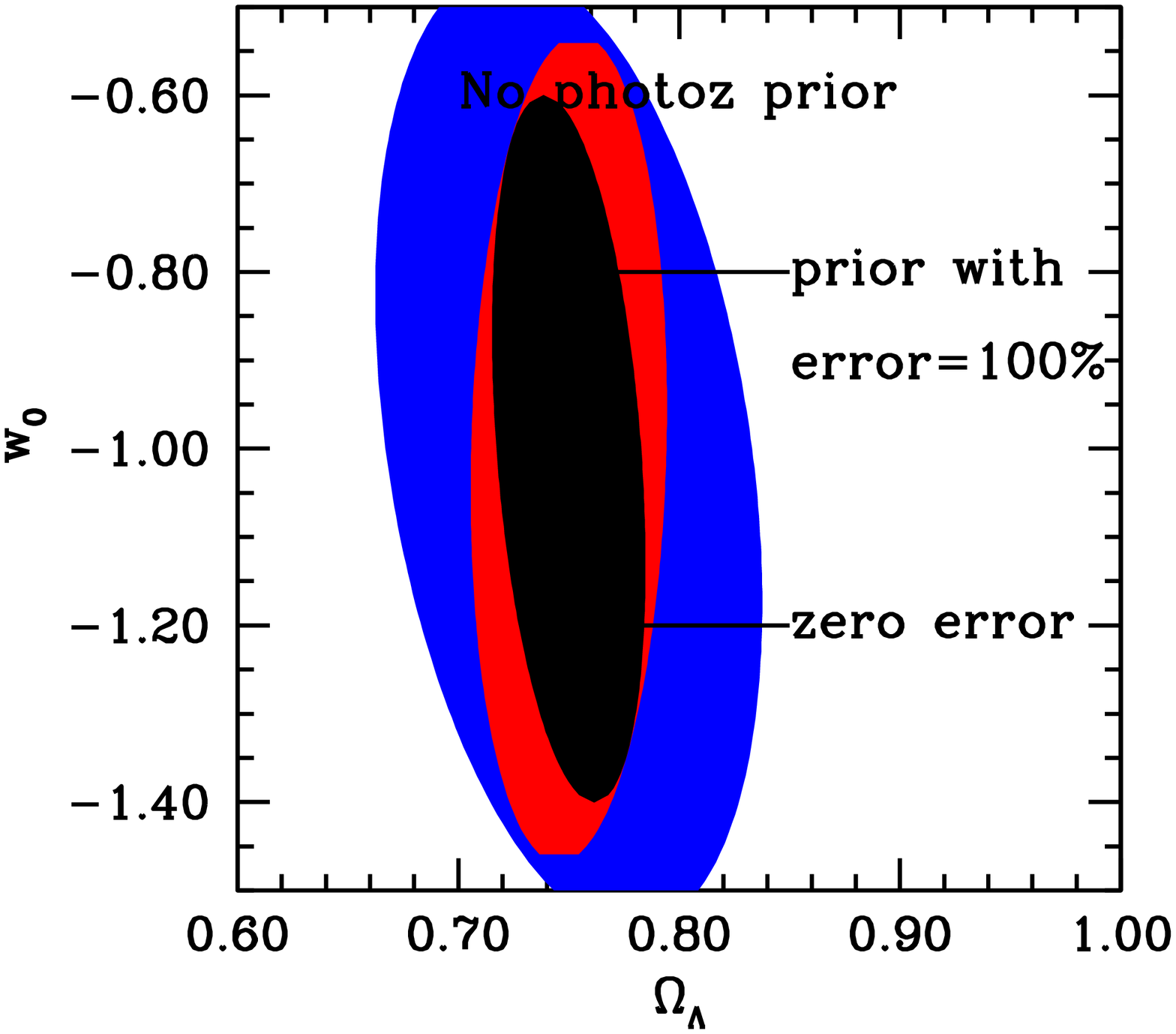}} 
      \resizebox{85mm}{!}{\includegraphics{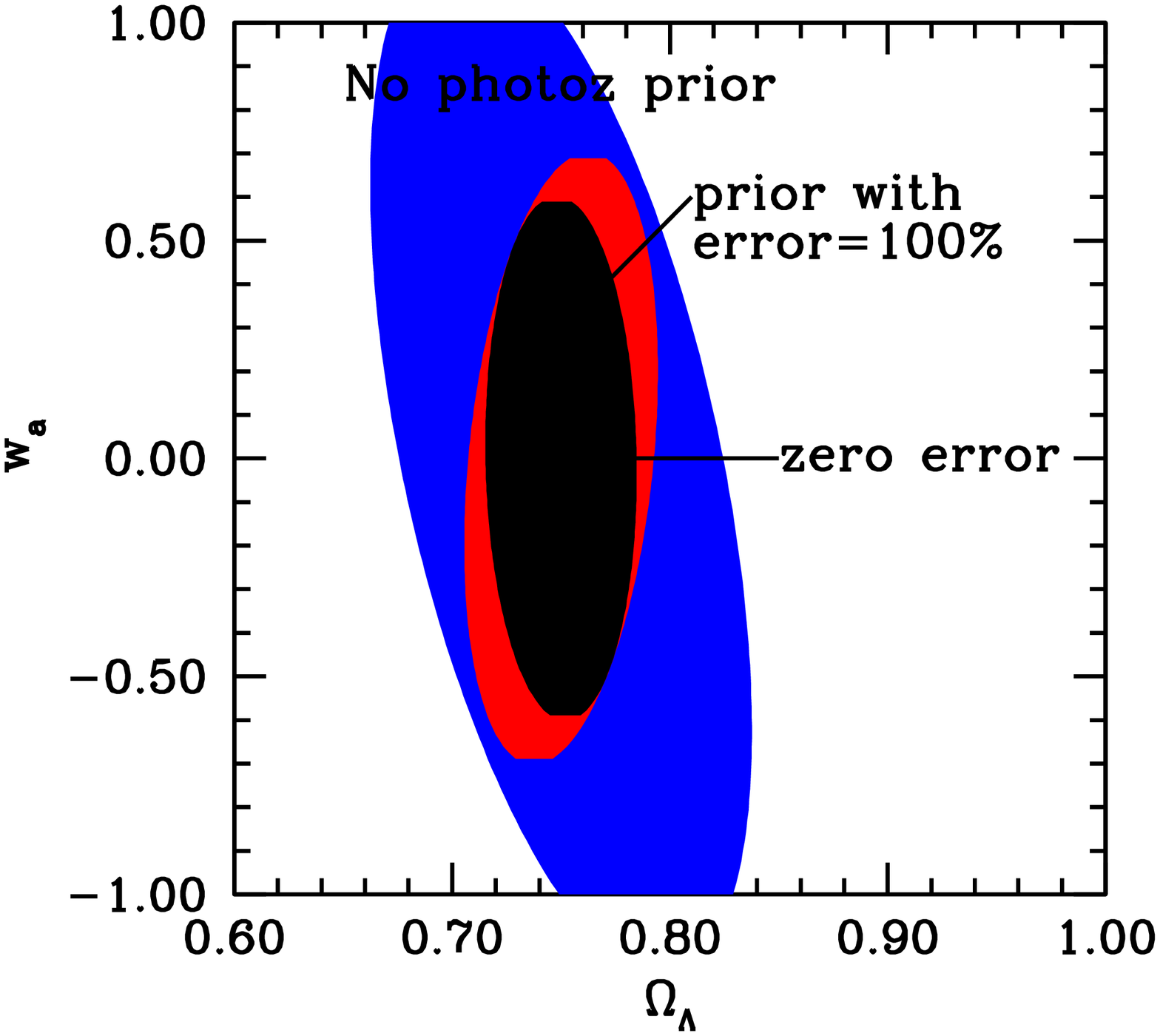}}
     \end{tabular}
    \caption{
Dark energy parameter constraints from mean pairwise SNIa velocities (in the absence of 
complementary cosmological probes), for the same survey as in Fig.~\ref{fig:StoN}.  
The left and the right panels show the 1$\sigma$ contour in the $w_0-\Omega_{\Lambda}$ 
and the $w_a-\Omega_{\Lambda}$ planes. Photometric redshift errors of $\sigma_z=0.01(1+z)$ 
are assumed. The black shaded region represents the case when the photo-z distribution is 
known accurately. The red (light shaded) region applies a prior such that the uncertainty 
in the photometric redshift error, $\delta \sigma_{z0}$, is equal to the value of $\sigma_{z0}$; 
this is highly conservative. The blue (gray) shaded region shows the constraint when we 
have no prior knowledge about the photo-z error distribution 
(e.g., zero supernovae with spectroscopic redshifts).
}
\label{fig: photoz_fisher}
\end{center}
\end{figure*}

Figure~\ref{fig: photoz_fisher} and Table~\ref{tab:photoz} show marginalized statistical constraints 
on dark energy parameters from mean pairwise velocity only, under three strongly different  
assumptions regarding the photometric redshift error.  
The blue (gray) and the black shaded regions show the two extreme cases.  
The blue shaded area shows the $1\sigma$ constraint when 
we assume no prior knowledge of the uncertainty in the photo-z error and allow 
$\sigma_{z0}$ to be determined from the same data used to constrain cosmology. 
The black region indicates the constraints when $\sigma_{z0}$ is known exactly.   
We emphasize that this does not mean that the photometric redshift is equal to the 
true redshift.  There is still a non-negligible dispersion in photometric redshifts 
in this case; however, we have assumed that the photometric redshift distribution 
is well understood, perhaps due to calibration with several 
thousand spectra \citep{zentner_bhattacharya09}.  The red (light shaded) region 
represents the case when the prior on $\sigma_{z0}$ is a Gaussian centered at 
the true value with sigma equal to its fiducial value, $\sigma_{z0}=0.01$.

Constraints on $w_0$, $w_a$ and $\Omega_\Lambda$ change by only about 10\% between 
the case where $\sigma_{z0}$ is uncertain at the 100\% level and one where we assume a perfectly-calibrated 
photometric redshift distribution.  This results from the fact that the mean pairwise velocity is 
proportional to the redshift difference between 
galaxies in a pair, but photometric redshift errors do not correlate with the velocity we are trying to measure.
Fig.~\ref{fig: photoz_fisher} shows that even weak prior knowledge of the photo-z distribution yields 
constraints comparable to a scenario where the photo-z error distribution is known exactly.

%%%%%%%%%%%%%%%%%%%%%%%%%%%%%%%%%%%%%%%%%%%%%%%%%%%%%%%%%%%%%%%%%%%%%%%%%%%%%%%%%%%

%%%%%%%%%%%%%%%%%%%%%%%%%%%%%%%%%
\begin{table}
\begin{center}
\begin{tabular}{| c | ccc |}
\hline
$\sigma(\sigma_{z0})$ & \ \ $\sigma(\Omega_\Lambda)$ & \ \ $\sigma(w_0)$ & \ \ $\sigma(w_a)$ \\ 
\hline
${\rm no\, prior}$ & \ \ 0.046 & \ \ 0.29 & \ \ 0.62\\
${\rm prior\, (100\%\, error\,  in\,  \sigma_{z0})}$ & \ \ 0.023 & \ \ 0.25 & \ \ 0.37\\
${\rm prior\, (zero\, error\,  in\,  \sigma_{z0})}$ & \ \ 0.018 & \ \ 0.21 & \ \ 0.31\\
\hline
\end{tabular}
\end{center}
\caption{
Impact of prior information about photo-z distributions on the dark energy parameter constraints derived 
from mean pairwise velocity statistics (calculated in the absence of complementary cosmological probes). 
The photometric redshift of an SN is assumed to be normally distributed about its true value with 
$\sigma_z=\sigma_{z0}(1+z)$; for our standard scenario we take $\sigma_{z0}=0.01$.  The fiducial 
values cosmology considered has $\Omega_\Lambda=0.75$, $w_0= -1$ and $w_a= 0$. 
We assume $\Omega_k=0$ and assume Gaussian priors with
standard error of 5\% on $h$ and $\Delta_\zeta$ and 1\% on $n_s$.}
\label{tab:photoz}
\end{table}
%%%%%%%%%%%%%%%%%%%%%%%%%%%%%%%%%%%%%%%%

%%%%%%%%%%%%%%%%%%%%%%%%%%%%%%%%%%%%%%%%%%%%%%%%%%%%%%%%%%%%%%%%%%%%%%%%%%%%%%%%%%
\section{Discussion and Future Prospects}
\label{sec: discuss}

With vastly increased numbers of Type Ia supernova detections on the horizon, a new statistical
probe of dark energy using supernova peculiar velocities will be possible. The Dark Energy Task
Force, when considering future supernova measurements, made the optimistic but reasonable
assumptions that individual supernovae will have a photometric redshift determined with
a standard error $0.01$ at redshift $z=0$, and a distance modulus determined with an error of 0.1.
If these levels are attained for the nominal $3\times 10^5$ SNeIa which will be detected in
a targeted supernova survey area by the LSST, the resulting dark energy constraints
from the mean pairwise velocities of these supernovae are interestingly good, comparable 
to projections for a variety of Stage IV techniques. In particular, pairwise peculiar velocities alone
give a slightly stronger dark energy constraint as the optimistic projection for an optical
baryon acoustic oscillation probe, and constraints which are towards the optimistic ends
of the galaxy cluster abundance and optical supernova Hubble diagram probes. 

%This technique is comparable to the Stage-IV techniques of galaxy cluster counts, baryon acoustic
%oscillations, and supernova luminosity distance for determining the dark energy parameters
%$w_0$ and $w_a$, while not as good as the optimistic projections for weak lensing
%in the radio band or from space. 

Having another independent method for constraining dark energy is invaluable, since
all of these measurements will likely be limited by systematic error control. Comparison
of inferred dark energy parameters from multiple independent experiments is even more
important than the combined statistical power of multiple measurements. In addition, 
the science return from mean pairwise velocity measurements comes essentially ``for free,''
as it uses the same data sets from which supernova luminosity distance measurements will 
be built.  

An extension of these observations which can significantly improve the strength of dark energy
constraints is the addition of spectroscopic redshifts. Surveys like Pan-STARRS and LSST will provide
only photometric redshifts, and the sheer number of objects they observe makes obtaining spectroscopic redshifts 
for even small subsets of the total objects a massive challenge. Given the numbers used in this paper, LSST will detect on the order of 100 new SNeIa per night of the survey. The supernovae themselves are transient and widely spread over the survey area, limiting the total number which can be observed simultaneously.  Obtaining immediate redshift follow-up for all of these objects would be a large logistical challenge, even if a dedicated telescope were available. 
An alternative is to obtain redshifts of host galaxies after their SNe have faded; this can be done much more efficiently,
as many host galaxies in a particular field of view could be targeted simultaneously
with multi-object spectrographs. Baryon acoustic oscillation observations, in particular, are pioneering the development of
very large fiber spectrographs which can obtain thousands of redshifts simultaneously.
The challenge for this strategy is that many of the host galaxies will be at redshifts above
0.5, and many of the host galaxies themselves are dim compared to their supernovae. Host galaxy
followup would require a large telescope and large amounts of observing time.  The spectrograph proposed for the BigBOSS survey, which has been designed to obtain high-throughput spectroscopy of 5000 galaxies at a time over a 7 square degree field of view using the 4m Mayall telescope at Kitt Peak and the Blanco telescope at CTIO \citep{schlegel09}, would be well suited for this task.  Such a spectroscopic survey of supernova hosts would be
a major undertaking, but could lead to highly competitive dark energy constraints and leverages instruments and data already planned for other purposes. 

Another possible avenue for improvement is better standardization of SNIa intrinsic luminosities.
Here our baseline assumption, along with the DETF, is that SNIa distance moduli will be known
with a standard error of around 0.1.  It is an open question whether we eventually will understand
the SNIa explosion mechanism in enough detail, and have sufficient observational
information, to model some portion of this scatter and reduce the effective random error. Magnification due
to gravitational lensing provides an additional source of scatter, which can be partly understood
due to its strongly non-Gaussian distribution, but for our nominal model survey we are not
limited by lensing scatter. The marginalized constraint on $w_a$, the most challenging dark energy
parameter to measure, can be improved by a factor of two if the scatter in the intrinsic
supernova distance modulus is halved. A mean pairwise velocity measurement for
$3\times 10^5$ supernovae with spectroscopic redshifts and an intrinsic distance modulus
scatter of 0.05 would constrain $w_a$ with a standard error of 0.08 using our set of current prior constraints.

The statistical power of any given dark energy measurement is only half of the story, as all of these
measurements are likely to be heavily dependent on systematic error control. 
Because of its nature as a differential measurement, the mean pairwise velocity technique offers 
favorable prospects for controlling systematic errors. Differential measurements have long been
exploited in measurements of the cosmic microwave background fluctuations precisely for their systematic
error advantages. In particular,
we have demonstrated that several obvious systematic error sources are not likely to dominate
the dark energy constraints. First, uncertainty about the level of scatter in photometric redshifts about
their true values has only a weak effect on dark energy constraints, and mild priors
obtainable from modest spectroscopic calibration efforts give results that are nearly the same
as exact knowledge of the photometric redshift scatter. We have not considered non-Gaussian
errors in photometric redshifts, but any scatter which is characterized at the levels of
the normal errors considered here is unlikely to induce any significantly larger systematic
errors. 

Second, a bias in the photometric redshift
distribution has very little effect on our constraints, as long as the bias varies slowly with
redshift, because a constant redshift bias doesn't affect the pairwise velocities. 
This is in marked contrast to both the supernova luminosity distance and weak lensing
techniques. Both of these widely discussed routes to dark energy constraints are very sensitive 
to photometric redshift biases \citep{huterer04,zentner_bhattacharya09}, where a redshift bias 
can mimic a shift in dark energy parameters. Third, a systematic error in distance modulus
due to unrecognized evolution in mean supernova luminosity with redshift will
be a small effect provided the magnitude of the error is within a factor of 3 of that
considered in the DETF report. This potential source of error can also be addressed by
testing the rich information in supernova spectra and time series at different redshifts for
any evidence of evolution in intrinsic supernova properties. While detailed modeling of 
potential systematic errors is required to understand any particular experiment, it is plausible
that the systematic errors associated with mean pairwise velocities will be substantially
less severe than other leading techniques for probing dark energy.

We also note that mean pairwise velocities can be used to constrain gravitational
explanations for the accelerating expansion of the universe. This technique has the
advantage of probing structure growth over a wide range in redshift, while also being
sensitive to the expansion rate; the comparison between these two quantities is the
key to constraining alternate gravity models \citep{jai08,hu07,lin09}. Pairwise velocities from 
a much smaller sample of galaxy clusters, with more precise velocities obtained via the kinematic
Sunyaev-Zeldovich effect, have already been shown to offer potentially interesting constraints
on modifications of gravity \citep{kb09}. 

The pairwise velocity statistic offers a particularly
simple route to a probe of modified gravity. In linear perturbation theory, the evolution
of the growth factor $D(a)$ is given to a very good approximation by $d\ln D/d\ln a = \Omega(a)^\gamma$,
where $\gamma$ is nearly a constant and takes the value $\gamma\approx 0.55$ for general
relativity \citep{linder05}; see Ref.~\citet{jus09} for a highly accurate approximation to $D(a)$. Other gravitation theories can have different values of $\gamma$; for example,
DGP gravity \citep{dgp00} has $\gamma=0.68$ \citep{linder_cahn08}.  Examining Eq.~(\ref{v12}), we see that the mean pairwise
velocity (on the left side) depends linearly on $d\ln D/d\ln a$, as well as $H(a)$, the (linear regime) galaxy bias factor, and correlation function information.  Other cosmological tests, such as the supernova Hubble diagram, will directly constrain $H(a)$, while correlation functions will be measurable directly from the data set used.  The linear clustering bias of host galaxies can be constrained in a number of ways; e.g., by direct comparison of galaxy correlation functions to the matter power spectrum derived from gravitational lensing; with galaxy three-point correlation functions \citep{mcbride10} or angular bispectra \citep{verde02_1}; or (if a large spectroscopic sample is available) by combining redshift-space distortions \citep{linder08,white08,percival08} with mean pairwise velocity statistics.
Assuming these other quantities will be measured with errors which are small compared to our velocity errors,
a measurement of $v(r,a)$ will provide an estimate of $d\ln D/d\ln a$ in several bins in $a$, which can then 
be used to constrain $\gamma$. 

If for each bin in $a$ we have independent measurements
of $v(r,a)$ in five radial bins with a signal-to-noise ratio of around 3 in each bin (see Fig.~\ref{fig:StoN}),
then the amplitude of the function $v(r,a)$ can be constrained with a fractional error of around
$0.33/\sqrt{5}$ or 0.16. Assuming this is the dominant error, the fractional error on $d\ln D/d\ln a$ is also around 0.16. 
By propagation of errors, the resulting error on $\gamma$ is then $0.16/\ln\Omega_m(a)$, and hence ranges from 0.15 to
0.45, depending on the redshift bin. This would give
approximately a 25\% to 80\% measurement of $\gamma$ in each redshift bin (assuming $\gamma$ takes its
general relativistic value), providing a significant constraint on many theoretical
alternatives to general relativity. With spectroscopic redshifts, these constraints would improve
by a factor of three due to the increase in signal-noise ratio in each angular bin; then the best redshift
bin alone might provide a 10\% measurement of $\gamma$, comparable to projected
constraints from weak lensing \citep{hearin_zentner09}. Prospects for constraining
modified gravity with a large supernova survey will be explored in more detail elsewhere. 

Dark energy is simultaneously one of the most important problems in physics today,
and one of the most elusive to address observationally. Mean pairwise velocities extracted from
a large survey of SNeIa can provide an important arrow in the dark energy quiver and
should be considered alongside any of the other methods now being actively
pursued. If simply piggybacked on existing plans for supernova luminosity distance tests,
pairwise velocities offer independent dark energy constraints which are competitive with other methods.
If augmented by spectroscopic redshift followup observations, pairwise velocities alone may provide important
constraints on dark energy, with constraints on $w_a$ of 0.1 or better.  Perhaps most importantly, this technique provides not only statistical power but potentially strong control of systematic errors.  Additionally, it allows tests of the nature of gravity which cannot be obtained using distance measurements alone.  We anticipate
that Type Ia supernova peculiar velocity statistics will be in the vanguard of dark energy constraints over
the coming years.

%%%%%%%%%%%%%%%%%%%%%%%%%%%%%%%%%%%%%%%%%%%%%%%%%%%%%%%%%%%%%%%%%%%%%%%%%%%%%%%%%%%%%%%%%%%%%%%%%%%%
\begin{acknowledgments}
The authors would like to thank Michael Wood-Vasey for useful 
discussions about the potential of imaging surveys to improve 
cosmological constraints from Type Ia supernovae and Daniel Holz 
for discussion about the lensing of supernovae. An anonymous referee
provided a number of helpful suggestions to clarify various points, and
prompted discovery of a factor-of-two mistake in computing statistical errors. 
SB was partly supported by the Mellon Predoctoral Fellowship at the 
University of Pittsburgh during this project and the LDRD program of Los Alamos National Lab.  
ARZ is funded by the University of Pittsburgh, 
the National Science Foundation through grant 
AST 0806367, and by the Department of Energy.
AK is supported by NSF grant AST 0807790.
JAN is funded by the University of Pittsburgh, 
the National Science Foundation through grant 
AST 0806732, and by the Department of Energy.
\end{acknowledgments}

\bibliography{paper5v4}

\end{document}